\documentclass[twocolumn]{aastex62}
\usepackage{amsmath,amstext}
\usepackage[T1]{fontenc}
\usepackage{apjfonts} 
\usepackage{auto-pst-pdf}
\usepackage[figure,figure*]{hypcap}

\newcommand{\ks}{\hbox{$K_s$}}

\newcommand{\lsim}{\lesssim}
\newcommand{\gsim}{\gtrsim}

\newcommand{\eg}{e.g.}

\newcommand{\Msol}{\hbox{$M_\odot$}}
\newcommand{\msol}{\hbox{$M_\odot$}}

\newcommand{\editone}[1]{{#1}}

\journalinfo{Accepted for publication in the \textit{Astrophysical Journal}}

\shorttitle{Environmental Evolution of the Galaxy Stellar Mass Function}
\shortauthors{Papovich et al.}

\begin{document}

\title{\uppercase{The Effects of Environment on the Evolution of the Galaxy Stellar Mass Function}}

\correspondingauthor{Casey Papovich}
\email{papovich@tamu.edu}

\author[0000-0001-7503-8482]{Casey Papovich}
\affiliation{Department of Physics and Astronomy, Texas A\&M University, College
Station, TX, 77843-4242 USA}
\affiliation{George P.\ and Cynthia Woods Mitchell Institute for
  Fundamental Physics and Astronomy, Texas A\&M University, College
  Station, TX, 77843-4242 USA}

\author[0000-0003-4032-2445]{Lalitwadee Kawinwanichakij}
\affiliation{Department of Physics and Astronomy, Texas A\&M University, College
Station, TX, 77843-4242 USA}
\affiliation{George P.\ and Cynthia Woods Mitchell Institute for
  Fundamental Physics and Astronomy, Texas A\&M University, College
  Station, TX, 77843-4242 USA}
\affiliation{LSSTC Data Science Fellow}  

\author[0000-0003-0341-8827]{Ryan F. Quadri}
\affiliation{Department of Physics and Astronomy, Texas A\&M University, College
Station, TX, 77843-4242 USA}
\affiliation{George P.\ and Cynthia Woods Mitchell Institute for
  Fundamental Physics and Astronomy, Texas A\&M University, College
  Station, TX, 77843-4242 USA}
\affiliation{Mitchell Astronomy Fellow}

\author{Karl Glazebrook}
\affiliation{Centre for Astrophysics and Supercomputing, Swinburne University of Technology, Hawthorn, Victoria 3122}

\author{Ivo Labb\'e}
\affiliation{Leiden Observatory, Leiden University, NL-2300 RA Leiden, The Netherlands}

\author[0000-0001-9208-2143]{Kim-Vy H. Tran}
\affiliation{School of Physics, University of New South Wales, NSW 2052, Australia}
\affiliation{Department of Physics and Astronomy, Texas A\&M University, College Station, TX, 77843-4242 USA}
\affiliation{George P.\ and Cynthia Woods Mitchell Institute for
  Fundamental Physics and Astronomy, Texas A\&M University, College
  Station, TX, 77843-4242 USA}

\author[0000-0001-6003-0541]{Ben Forrest}
\affiliation{Department of Physics and Astronomy, Texas A\&M University, College
Station, TX, 77843-4242 USA}
\affiliation{George P.\ and Cynthia Woods Mitchell Institute for
  Fundamental Physics and Astronomy, Texas A\&M University, College
  Station, TX, 77843-4242 USA}

\author{Glenn G. Kacprzak}
\affiliation{Centre for Astrophysics and Supercomputing, Swinburne University of Technology, Hawthorn, Victoria 3122}

\author[0000-0001-5185-9876]{Lee~R.~Spitler }
\affiliation{Research Centre for Astronomy, Astrophysics \& Astrophotonics, Macquarie University, Sydney, NSW 2109, Australia}
\affiliation{Department of Physics \& Astronomy, Macquarie University, Sydney, NSW 2109, Australia}
\affiliation{Australian Astronomical Observatories, 105 Delhi Rd.,
Sydney NSW 2113, Australia}

\author{Caroline M. S. Straatman}
\affiliation{Max-Planck Institut f\"ur Astronomie, K\"onigstuhl 17, D$-$69117, Heidelberg, Germany}

\author[0000-0003-2008-1752]{Adam R. Tomczak}
\affiliation{Department of Physics, University of California, Davis, One Shields Ave., Davis, CA 95616, USA}


\begin{abstract} 
We study the effects of galaxy environment on the evolution of the
stellar--mass function (SMF) over $0.2<z<2.0$ using the FourStar Galaxy
Evolution (ZFOURGE) survey and NEWFIRM Medium--Band Survey (NMBS) down
to the stellar--mass completeness limit, $\log M_\ast/M_\odot>9.0$
(9.5) at $z=1.0$ (2.0).  We compare the SMFs for quiescent and
star--forming galaxies in the highest and lowest environments
using a density estimator based on the distance to the galaxies'
third--nearest neighbors.  For star--forming galaxies, at all redshifts
there are only minor differences with environment in the shape of the
SMF.  For quiescent galaxies, the SMF in the lowest densities shows no
evolution with redshift, other than an overall increase in number
density ($\phi^\ast$) with time.  This suggests that the stellar--mass
dependence of quenching in relatively isolated galaxies is both
universal and does not evolve strongly.  While at $z\gsim1.5$ the
SMF of quiescent galaxies is indistinguishable in the highest and
lowest densities, at lower redshifts it shows a rapidly increasing
number density of lower--mass galaxies, $\log
M_\ast/M_\odot\simeq9-10$.  We argue this evolution can account for
all the redshift evolution in the shape of the \textit{total}
quiescent--galaxy SMF.  This evolution in the quiescent--galaxy SMF at
higher redshift ($z>1$) requires an environmental--quenching efficiency
that decreases with decreasing stellar mass at $0.5<z<1.5$ or it
would overproduce the number of lower--mass quiescent galaxies in
denser environments.   This requires a dominant environment process
such as starvation combined with rapid gas depletion and ejection at
$z>0.5-1.0$ for galaxies in our mass range.  The efficiency of this
process decreases with redshift allowing other processes (such as
galaxy interactions and ram--pressure stripping) to become more
important at later times, $z<0.5$.   
\end{abstract}

\section{Introduction} \label{section:intro}

How galaxies quench their star-formation depends on the interplay
between gas accretion, gas cooling, and the strength and timescales of
feedback \citep[see \eg,][]{some15,feld17}.  In the nearby and distant
Universe, studies show the rate of quenching and the quiescent--galaxy
fraction are correlated with both increasing stellar mass and local
environment (galaxy density)
\editone{\citep{baldry06,peng10,vulc12,kovac14,balogh16,davies16,darv17,kawi17,nant17}}, implying that
there are processes that affect galaxy quenching that depend on galaxy
total mass and on galaxy environment.  In the local Universe,
these effects may be separable \citep{baldry06,peng10}, but in the
more distant Universe, the evidence suggests otherwise
\citep[see][]{kovac14,balogh16,kawi17}.  It is important to quantify
this evolution in the strength of mass quenching and environmental
quenching because these constrain the mechanisms and timescales of
the physical processes themselves.  

If the strength of quenching depends on galaxy redshift, stellar mass,
and environment, then this should be visible in the differential
evolution of the stellar mass functions (SMFs) of galaxies.   
Observations of the galaxy SMF in the nearby Universe with the
\textit{Sloan Digital Sky Survey} \citep[SDSS; at $z\sim
0.085$,][]{baldry06,peng10} show significant differences for star-forming and
quiescent galaxies as a function of environment, including a 
steeper low-mass slope for quiescent galaxies in denser environments.

In the more distant Universe, measurements of the dependence of SMF
on environment have yet to reach consensus.  Most studies
compare the SMF in massive groups and clusters to the field over $0 <
z < 2$.  Some find no evidence for significant differences \citep[once
the brightest cluster galaxy is excluded;][although see,
\citealt{rudn12,tomc17}]{vulc12,vulc13,andr13,vanderburg13,nant16}.  Other
studies find differences at
relatively lower redshift ($z \lsim 0.8$), that disappear by $z \sim
1$ \citep{bolz10,giod12,mok13,ethe17}, at least for galaxies more
massive than ``$M^\ast$'' (the characteristic mass of the SMF).  This
is consistent with other studies of the SMF in denser environments
such as in high-redshift ($0.8 < z < 1$) groups
\citep[\eg,][]{mok13,balogh16,tomc17} and high redshift ($1 < z <
1.5$) clusters \citep{andr13,vanderburg13,nant16,tomc17}, where some
have found a deficit of low-mass quiescent galaxies (and low-mass
star-forming galaxies) compared to mass-matched field samples.

One complication is that many of the studies of groups and clusters
compare to ``field'' surveys that include rich groups and clusters.
For example, the 1.6 deg$^2$ UltraVISTA survey of the COSMOS field
\citep{muzzin13b} is a frequently-referenced field sample, but it
covers a range of environmental densities and includes rich groups
\citep[or even poor clusters, e.g.,][]{giod12}.  If environmental
quenching or ``preprocessing'' occurs in low-mass group environments,
then care must be made to ensure that the effects of environment do
not bias the results from field samples.  Studies that separate
galaxies in field surveys into samples of high and low--density
regions find stronger differences in the SMF at least to $z < 1.3$
\citep[\eg,][]{tomc17}, with no measurable difference at higher
redshift \citep[although studies report tentative evidence that the
low-mass slope of the SMF is steeper for galaxies in higher density
environments out to $z \sim 1.5$;][]{mort15}.   

Because we have yet to obtain good constraints on the evolution of the
SMF of star-forming and quiescent galaxies in different environments,
we have yet to constrain the dominant environmental quenching
processes, and to determine if these processes change with time.
 Here we study the
evolution of the galaxy SMF as a function of environment,
star-formation activity, and redshift over $0.2 < z < 2.0$ using two,
homogeneous datasets that combine differing depth and area, the
photometric-redshift FourStar Galaxy Evolution (ZFOURGE) survey
\citep{stra16} and the NEWFIRM Medium-Band Survey
\citep[NMBS]{whit11}.   \editone{ Both surveys provide very accurate
photometric redshifts, 
%
%
able to resolve structures on scales of $<$4000 km s$^{-1}$.}  This
allows for the identification of rich galaxy overdensities
\citep[\eg,][]{spit12,forr17} and minimizes inaccuracies associated
with environmental measures derived from photometric redshift surveys
with larger redshift errors \citep[see, e.g.,][]{cooper05,mala16}.
Recently, \citet{kawi17} used these ZFOURGE data to measure the
environmental quenching efficiency.  They quantified the excess
quenching as a function of increasing galaxy overdensity (environment)
and showed this must \textit{decrease} with stellar mass for galaxies
at high redshifts to account for the relatively low fraction of
quenched galaxies in \textit{any} environment at high redshift.
Here, we use these data to study how and when the environment affects
the \editone{build--up of the number density of low-mass quiescent
galaxies.}

Following \citet{kawi17}, we measure the local galaxy density as a
proxy for environment using a Bayesian-motivated measure of the
distance to the third--nearest neighbor (3NN), introduced by
\citet{ivezic05} \citep[see also][]{cowan08}.  \citet{muld12} find
that the $N$th--nearest neighbor technique best probes the local
environment on scales internal to galaxy halos for lower values of
$N$. \citeauthor{kawi17} demonstrate that quantifying
galaxy density with 3NN recovers physical structures and robustly
identifies galaxies in the highest and lowest regions of the density
distribution using the same datasets used here. \editone{In addition,
we argue that using a lower value of $N$ (here $N$=3) to select
overdensities is appropriate to this work as it will allow us to
identify both rich groups and clusters \citep[see,
e.g.,][]{muld12,shat13}.    This allows us to
identify a more complete range of structures at higher redshift that
will collapse to cluster--sized objects at $z=0$.   For example, using
a set of simulations, \citet{shat13} show that selecting overdensities
with the $N$th nearest neighbor (for $N \leq 10$)  identifies a range of
overdensity covering a 2--3 dex range at $z=2$ that become very
overdense at $z=0$  (with typical $z=0$ halo masses $\log (M_h / M_\odot)
\sim 14.5$, with a spread of about $0.5$~dex).   }  In the present work,
we select galaxies in the highest and lowest density quartiles based
on the 3NN density measurements, and we compare the evolution of the
SMF for star-forming and quiescent galaxies in these regions.  

Following literature conventions \citep[see,
e.g.,][]{peng10,davies16,kawi17}, we refer to two kinds of quenching,
that which correlates with galaxy stellar mass (``mass quenching''),
and that which correlates with galaxy overdensity (``environmental
quenching'').   Here,  ``mass quenching'' is any process that acts
internally to a galaxy, and ``environmental quenching'' is any process
that is related to the local environment.    These may both be
manifestations of the same physics: e.g., they may be related to
quenching due to processes associated with galaxy halo mass
\citep[``halo quenching'', e.g.,][]{dekel06a,catt08}.  Mass quenching
would then correspond to the quenching of central galaxies, where
their stellar mass scales approximately with halo mass.  Environmental
quenching would result from processes that act as galaxies become
satellites, and also be related to the halo mass of the
\textit{central} galaxy.  Nevertheless, in this work we differentiate
quenching that correlates with stellar mass (mass quenching) from
quenching that correlates with overdensity (environmental quenching)
as this ties the measurements as closely as possible to observables. 

The outline of this paper is as follows.  \S~\ref{section:data}
discusses the datasets and sample selection.  \S~\ref{section:env}
describes our estimate of the galaxy environment.
\S~\ref{section:smf} presents the galaxy SMFs for these samples,
including the SMF for quiescent and star-forming galaxies as a
function of redshift and environment.   \S~\ref{section:discussion}
discusses the implications for environmental quenching processes, and
how these impact the evolution of the shape of the galaxy SMF.
Throughout, the majority of the emphasis is on the  evolution of the
SMF of quiescent galaxies, and  Appendix~\ref{section:appendix}
discusses the (lack of) evolution in the shape of the SMF for
star-forming galaxies.  

Throughout we use a cosmology with $\Omega_M=0.3$,
$\Omega_\Lambda=0.7$, and $h$=0.7, where $H_0 = 100$ $h$ km s$^{-1}$
Mpc$^{-1}$ consistent with the recent constraints from Planck
\citep{planck15} and the local distance scale of \cite{riess16}.   We
also assume a universal Chabrier IMF for the derivation of galaxy
stellar masses.  All magnitudes are in ``absolute bolometric'' (AB)
units \citep{oke83}. 

\section{Data and Sample Selection}\label{section:data} 

We use data from the ZFOURGE and NMBS public catalogs
\citep{whit11,stra16}.  For ZFOURGE, we used the v3.4
$\ks$-band--selected catalog reaching $\ks = 25.5-26.0$~mag (80\%
completeness) in the three fields (CDF--S, COSMOS, and UDS)  covering
a total of about 300 arcmin$^2$ \citep{stra16}.   For NMBS, we use the
$\ks$-band catalogs for the Cosmic Evolution Survey (COSMOS) and
All-wavelength Extended Groth strip International Survey (AEGIS)
fields, which cover a larger area ($\approx$1500 arcmin$^2$) but to
shallower depth, $\ks = 22.1-22.5$~mag \citep[][90\%
completeness]{whit11}.   Both the ZFOURGE and NMBS catalogs include
colors measured from medium-band imaging, \editone{which provide
accurate photometric redshifts, $\Delta z / (1+z) \simeq 1-2$\%}, and
rest-frame $(U-V)_0$ and $(V-J)_0$ colors.   \editone{From the
photometric catalogs and redshifts, both ZFOURGE and NMBS provide
stellar masses, where the ZFOURGE dataset achieves limiting stellar
masses of $\log M_\ast/M_\odot$$>$9.5 at $z$=2.0, reaching to low-mass
galaxies (1 dex below $M^\ast$).  We refer the reader to
\citet{whit11} and \citet{stra16} for details.}   

We classify galaxies as
quiescent or star--forming based on their rest-frame $(U-V)_0$ and
$(V-J)_0$ colors, where quiescent galaxies are red in $(U-V)_0$ and
relatively blue in $(V-J)_0$ compared to mass-matched star-forming
galaxies \citep[\eg,][]{labbe05,wuyts07,will09,whit11,papo15}.  We
select quiescent galaxies that satisfy, 
\begin{eqnarray}
(U-V)_0\ &\ge\ & 1.3\ \mathrm{mag},\ \nonumber\\
(V-J)_0\ &\leq\ & 1.6\ \mathrm{mag},\ \mathrm{and}\  \\
(U-V)_0\ &\ge\ & A\times (V-J)_0 + B, \nonumber
\end{eqnarray} 
where star-forming galaxies lie outside this region.
Here, we use $(A,B) = (1.2, 0.20)$ and (0.88, 0.65) for ZFOURGE
\citep{kawi16} and NMBS \citep{whit11}, respectively.   The values for
$A$ and $B$ depend on the specifics of each dataset, and exhibit a
possible dependence on redshift.  The slope, $A$, is chosen to run
parallel to the sequence of red galaxies, and the intercept, $B$, is chosen to
place the dividing line at the local minimum between the sequences of
quiescent and star-forming galaxies.  This has the effect of
minimizing uncertainties in the sample selection  arising from scatter
in the galaxies' rest-frame colors \citep[see discussion
in][]{kawi16}.  While different color selection limits change the
sample somewhat, this is not a dominant effect as \editone{it affects
relatively few galaxies} for sensible choices of $A$ and $B$.
\editone{For example, adopting the NMBS values for $A$ and $B$ would
  decrease the number of quiescent galaxies in ZFOURGE by $\simeq$6\%
  (and vice versa).   This is smaller than the expected fractional error
  from other effects, such as stellar mass uncertainties \citep[$\sim$10\%
  statistically for quiescent galaxies, although systematics can be
  factors of $\sim$2, \eg,][]{papo06a,marc09,muzzin09b,bram11} and
  rest-frame color uncertainties \citep[$\sim$0.1~mag, \eg,][]{whit11}.}


\section{Estimate of Environment}\label{section:env}

We use the local surface density of galaxies, $\Sigma$, as an
estimator of the local environment of each galaxy, as derived by
\citet{kawi17}.   This approach is based on the projected distance to
the $N$th--nearest neighbor, $d_N$, where the local galaxy surface
density is then $\Sigma_N = N (\pi d_N^2)^{-1}$.  As discussed by
\citet{ivezic05}, the precision of this approach can be improved (by
a factor of order 2) using a Bayesian estimator that includes the
distances of all neighbors up to the $N$th--nearest neighbor.
\citet{kawi17} use this estimator based on work of \citet{cowan08},
where the local surface density of galaxies is 
\begin{equation}
\Sigma_N^\prime = C \frac{N}{\Sigma_{i=1}^{N} d^2_i},
\end{equation}
where $d_i$ is the distance to the $i$th--nearest neighbor and $C$ is
chosen such that $\Sigma_N^\prime$ matches the density of a uniform
grid of points.   As in \citet{kawi17}, we use $N$=3, which provides
the best compromise between recovering the highest-- and lowest--
density regions and minimizing line-of-sight projections. Using this
(relatively low) value,  $N$=3, makes our measurement more sensitive
to halos of $L^\ast$--sized galaxies and groups ($\log M_h/\msol \gsim
12.5$), with relatively low contamination of galaxies in poor
environments of lower-mass halos with few satellites
\citep[\eg,][]{muld12}.  This is advantageous as many of the
environmental effects are expected to manifest in such regions
\citep[see \S~\ref{section:intro}, and discussion in][]{kawi17}. 

For this work we contrast the properties of galaxies in the top
quartile of the density distribution ($D4$, i.e., the highest density
quartile) with those in the bottom quartile of the density
distribution ($D1$, i.e., the lowest density quartile), similar to the
work of \citet{peng10}.  As in \citet{kawi17}, we determined these
quartiles using a non-parametric quantile regression method
\citep[specifically, the COnstrained B-Splines (COBS) linear
regression method,][]{ng07,feig12} applied to the ZFOURGE and NMBS
data.  For the ZFOURGE stellar-mass limit and our choice of redshift
binning, the surface densities dividing the quartiles are roughly
constant with redshift \citep[using a redshift window of $\Delta z =
0.05(1+z_\mathrm{phot})$ when identifying neighboring galaxies; see
Figure 2 of][]{kawi17}.  In this redshift interval, to our stellar
mass limit, the median surface density of galaxies is approximately 30
arcmin$^{-2}$, independent of redshift \citep{kawi17}.   The limiting
surface density for the highest ($D4$) density quartile is roughly
$\Sigma_3^\prime$$>$ 43 galaxies arcmin$^{-2}$, and for the lowest
density quartile ($D1$) it is roughly $\Sigma_3^\prime$$<$13 galaxies
arcmin$^{-2}$, both mostly independent of redshift
\citep[see][]{kawi17}.

\begin{figure*}
\epsscale{1.15}
\plotone{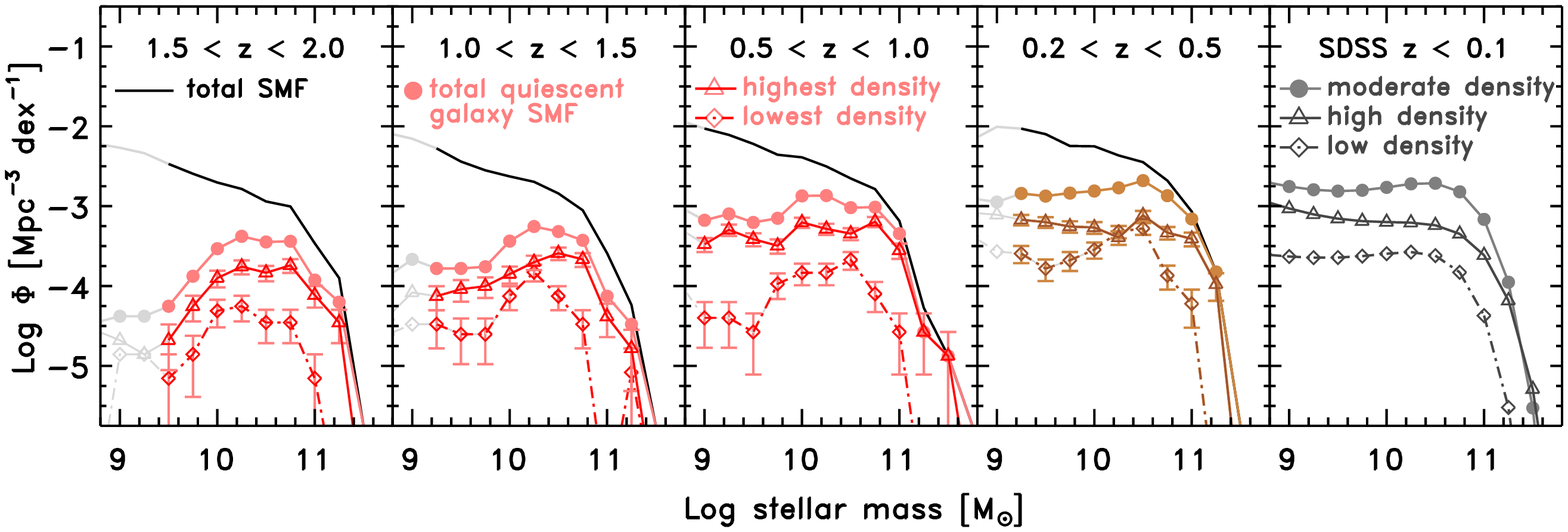}
\plotone{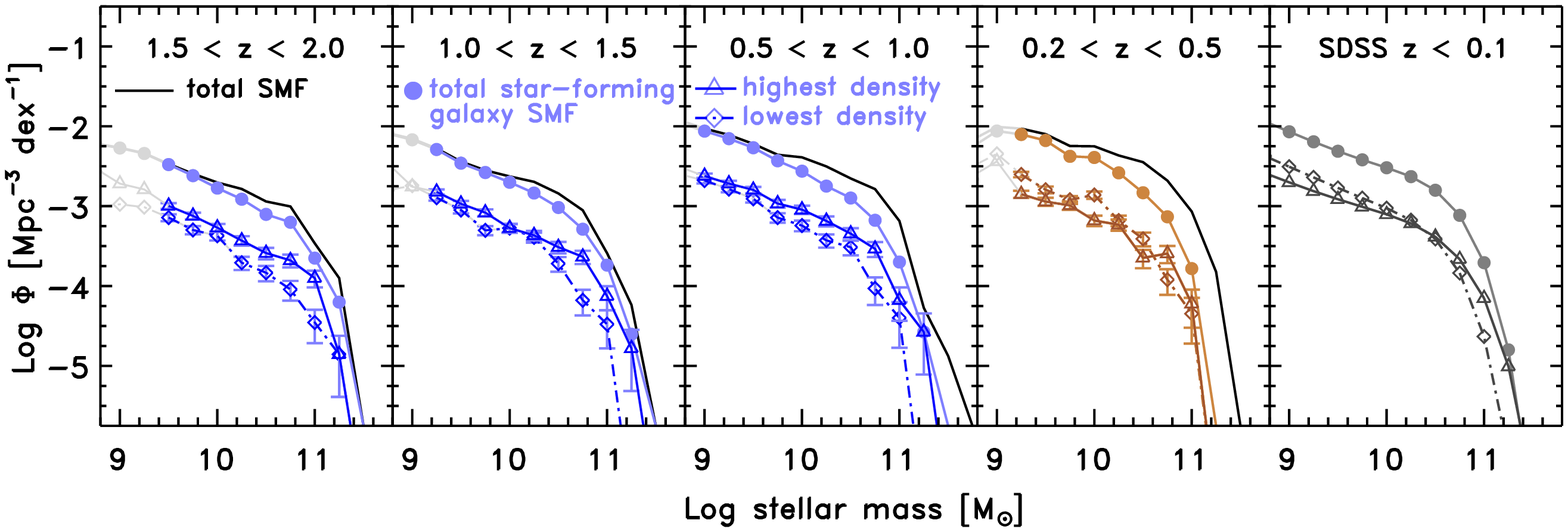}
\epsscale{1.}
\caption{Evolution of the Stellar Mass Function (SMF) for quiescent
and star-forming galaxies from the ZFOURGE survey ($0.5 < z < 2.0$)
and NMBS ($0.2 < z < 0.5$).   All panels show the total galaxy SMF
(solid black line) as a function of redshift, as labeled.   The top
row of panels shows the evolution of the SMF for all quiescent galaxies, and
for quiescent galaxies in the highest density quartile ($D4$) and
lowest density quartile ($D1$), as labeled.   The bottom row of panels shows
the evolution of the SMF for all star-forming galaxies and for
star-forming galaxies in $D4$ and $D1$, as labeled.  Light-gray shaded
points and lines show where the data fall below the stellar mass
completeness for the redshifts of each panel.  Error bars correspond
only to Poissonian uncertainties using the number of galaxies in each
data point.   The right-most panels in each row show SMFs from SDSS
at $z < 0.1$ derived in high densities, moderate densities, and low
densities \citep{baldry06}, scaled to match the normalization of our
SMFs derived from the NMBS at $0.2 < z < 0.5$ (see text).   The SMF for quiescent
galaxies shows a dependence on density while there is no such strong
dependence for the SMF of star-forming galaxies.  \label{fig:smf}}
\end{figure*}

\begin{figure*}
\epsscale{1.05}
\plotone{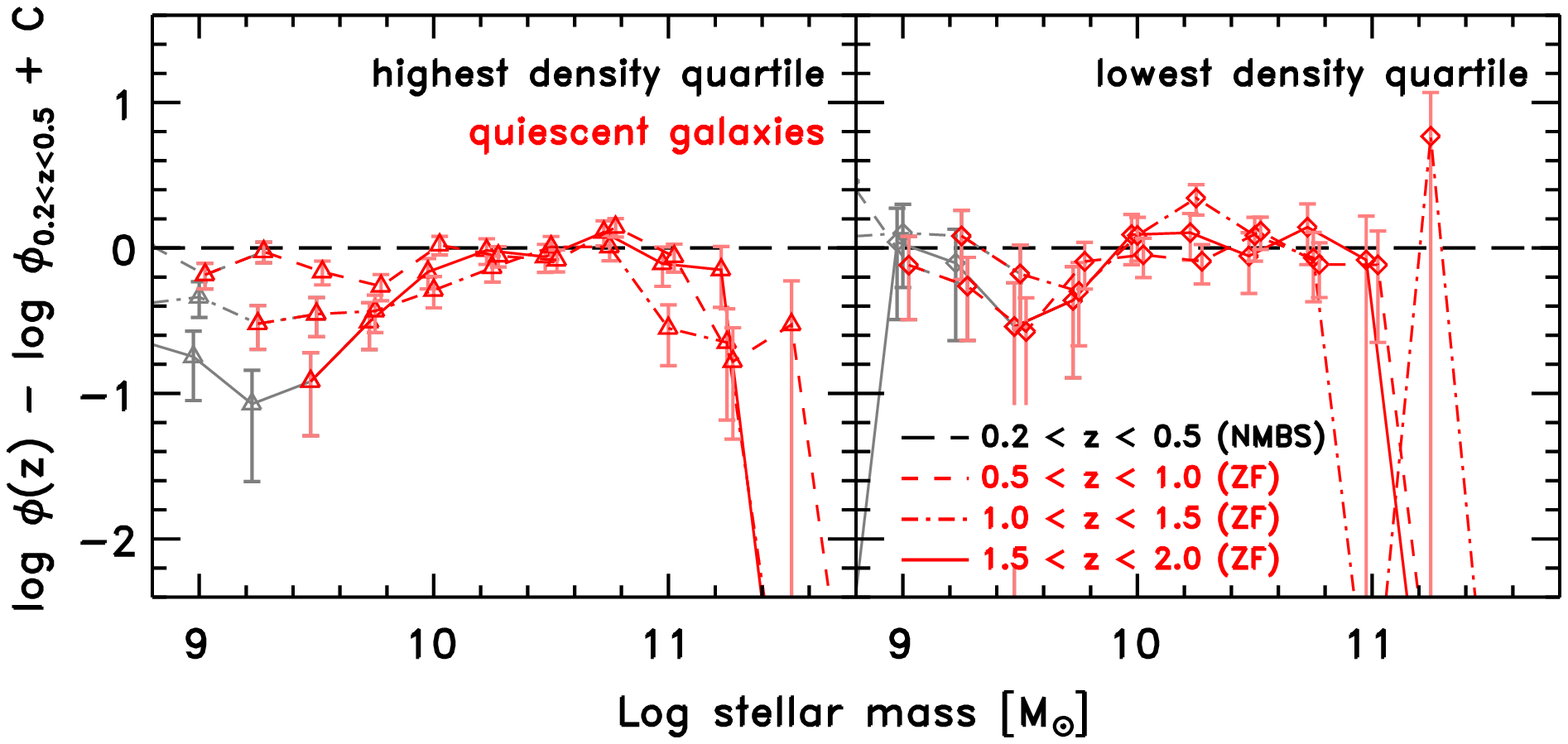}
\epsscale{1.0}
\caption{Relative evolution of the SMF for quiescent galaxies as a
function of environment.  The left panel shows the ratio of the SMF
for quiescent galaxies in the highest density quartile ($D4$) in each
redshift bin to the SMF for quiescent galaxies in the highest density
quartile at $0.2 < z < 0.5$ (the constant $C$ acts to normalize each
SMF to the same number density at $\log M_\ast/M_\odot$=10.6).    The
right panel shows the same for quiescent galaxies in the lowest
density quartile ($D1$)   The gray-shaded points and lines show data
below the stellar mass completeness limit.  There is no evidence that
the shape of the SMF evolves for \editone{quiescent galaxies in the
lowest--density quartile}.   In contrast, in the highest density
quartile, there is rapid redshift evolution in the SMF, particularly
in the relative number density of low-mass quiescent
galaxies.\label{fig:smf_ratio}}
\end{figure*}

\section{The Dependence of the Stellar Mass Function on
Environment}\label{section:smf}

To construct the SMFs, we sub-divided the galaxies into bins of
redshift and  star-formation activity (quiescent and star-forming),
and as a function of environment using the lowest--density ($D1$) and
highest--density quartiles ($D4$) as described above
(\S~\ref{section:data}).  The SMFs are then
\begin{equation}
\phi(m) = \frac{1}{\Delta m}\ \sum^N_{i=1}\frac{1}{V_c}, 
\end{equation}
where $m = \log M_\ast/\Msol$ is the base-10 logarithm of the stellar
mass, and the sum is over all $N$ galaxies with (log) stellar mass
between $m$ and $m+\Delta m$.  $V_c$ is the comoving volume and
depends on the redshift bounds that mark the sample, and geometry of
the survey.   Figure~\ref{fig:smf} shows these SMFs compared to the
total galaxy SMFs from ZFOURGE in bins of redshift, from $z=0.5$ to
2.0.\footnote{We do not show results for the middle quartiles (i.e.,
$D2$ and $D3$) because they include a mix of galaxies in both high and
low overdenstities.  \citet{kawi17} showed that data with the
photometric accuracy of ZFOURGE reliably recover galaxies in the
highest ($D4$) and lowest ($D1$) overdensities, but $D2$ and $D3$
may suffer higher contamination  because of redshift errors and
line-of-sight projections.  For this reason we focus only on the
results of the highest and lowest quartiles here, but we note that our
inspection of the SMFs for the $D2$ and $D3$ quartiles shows them to
lie between those of $D1$ and $D4$, and they do not change the
conclusions here.}  Note that we have not corrected any of the SMFs
for incompleteness in stellar mass (in which case  $V_c$ is the same
for all galaxies in a given redshift bin), but we denote the stellar
masses where the SMFs become incomplete using values derived in
\citet{kawi17}.  We furthermore have not corrected the environmental
measures for ``edge effects'' \citep[see][]{kawi17}. 

Figure~\ref{fig:smf} shows the SMFs we derived from ZFOURGE at $0.5 <
z < 2.0$, and NMBS $0.2 < z < 0.5$ for quiescent and star-forming
galaxies as a function of overdensity.  The figure also shows SMFs
from SDSS at $z < 0.1$ taken from \citet{baldry06} for red-- and
blue--sequence galaxies (akin to the quiescent and star-forming
galaxies studied here) in regions of high ($\log \Sigma = 1.0$),
moderate ($\log \Sigma = 0$), and low ($\log \Sigma = -1.1$)
overdensity scaled to match the number density of our SMFs in NMBS at
$0.2 < z < 0.5$.

From Figure~\ref{fig:smf} we draw our first two conclusions regarding
the dependence of the SMF with environment and star-formation
activity.  First, for star-forming galaxies there is little
environmental dependence in the shape of the SMF.  At all redshifts
from $z\sim 0$ to 2 the shape and normalization of the SMF of
star-forming galaxies in $D1$ and $D4$ are approximately
indistinguishable.  However, at $z > 0.5$ the SMF of star-forming
galaxies in the highest densities does show an excess in the number
density of $\log M_\ast/M_\odot \simeq 10.5$ galaxies compared to that
at lowest densities.  This may indicate the propensity of more massive
star-forming galaxies to be found in richer environments at higher
redshifts \citep[\eg,][]{quadri12,kawi17}.  This excess declines (or
even reverses) at $z < 0.5$ \citep[see also][]{peng10}. 

Second, for quiescent galaxies the shape of the SMF depends strongly
on environment and redshift.  At all redshifts, the normalization of
the SMF is higher in denser environments.  The shape of the SMF also
evolves with redshift: there is a rapid increase in the number density
of quiescent galaxies in higher density environments with decreasing
redshift, particularly at lower stellar masses.  These differences
become more pronounced at lower redshifts ($z \lsim 1$), and the trend
continues from our ZFOURGE dataset at $z>0.5$, to
the NMBS dataset over $0.2 < z < 0.5$, and down to $z < 0.1$ in SDSS
\citep[see also][]{baldry06,peng10}.  
%

Upon further scrutiny, the shape of the SMF of quiescent galaxies in
the lowest density ($D1$) quartile shows \textit{no} evidence of any
evolution  over the entire redshift range, $0.2 < z < 2.0$.
Figure~\ref{fig:smf_ratio} shows the relative differences in the
quiescent galaxies SMF in $D1$ and $D4$, compared to that in the NMBS
data at $0.2 < z < 0.5$ (where we have normalized each SMF to have the
same number density at $\log M_\ast/M_\odot$=10.6).  The right panel
of Figure~\ref{fig:smf_ratio} shows that the SMFs of quiescent
galaxies in the lowest density quartiles are consistent with no
evolution in redshift (within the uncertainties).  This suggests that
in the lowest densities, the (stellar) mass dependence of quenching in
relatively isolated galaxies does not evolve with time, at  least down
to our mass-completeness limit \citep[$\log M_\ast/\msol > 9.0-9.5$,
consistent with the suggestion by][]{peng12}.    This is not to say
that there is no environmental quenching in such systems.  For
example, even relatively isolated $L^\ast$ galaxies (including the
Milky Way Galaxy) show high fractions of quenched satellites,
particularly at lower stellar masses \citep[$\log M_\ast/M_\odot
\lesssim 8.5$,][]{geha12,slater14,wetzel15,davies16,geha17}, which
show environmental processes are at work.      Rather our result
implies that the stellar--mass dependence of quenching (either mass or
environmental) does not evolve in these low--density environments over
the range of galaxy stellar masses considered here. 

 For quiescent galaxies in the highest density environments,
Figure~\ref{fig:smf_ratio} shows that the shape of the SMF evolves
strongly with mass and redshift.  The effect is most pronounced at low
stellar masses $9 < \log M_\ast/\msol < 10.3$, where there is rapid
evolution from $1.5 < z < 2.5$ to $0.5 < z < 1.0$, with less change in
the SMF shape from $0.5 < z < 1.0$ to $0.2 < z < 0.5$.  This is
important as there is nearly twice as much cosmic time from $z=1.0$ to
0.2 (5.4 Gyr) as from $z=2.5$ to 1.0 (3.2 Gyr), \editone{implying
environmental quenching must act on timescales much shorter than this}
\citep{quadri12,balogh16,guo17}.     In contrast, for star-forming
galaxies, there is no evidence for evolution in the shape of the SMF,
see Appendix~\ref{section:appendix}.

\begin{deluxetable*}{lcccc}
\tablecaption{\editone{Results of Statistical Tests Comparing Stellar Mass
  Distributions of Galaxy Subsamples\label{table:pvalues}}}
\tablecolumns{5}
\tablewidth{0pc}
\tablehead{
\colhead{Galaxy Population} &
\colhead{Density Quartile} &
\colhead{Redshift Ranges Compared} & 
\colhead{KS test $p$-value}  & 
\colhead{MWW test $p$--value} \\
\colhead{(1)} & 
\colhead{(2)} & 
\colhead{(3)} & 
\colhead{(4)} & 
\colhead{(5)} }
\startdata
Quiescent Galaxies & $D4$ & $(0.5 < z < 1.0)$~~~to~~~$(1.0 < z < 1.5)$ & $6.6
\times 10^{-5}$ & $8.3\times 10^{-4}$ \\
 & $D4$ & $(0.5 < z < 1.0)$~~~to~~~$(1.5 < z < 2.0)$ & $1.0
\times 10^{-5}$ & $4.6\times 10^{-6}$ \\ 
 & $D4$ & $(1.0 < z < 1.5)$~~~to~~~$(1.5 < z < 2.0)$ & 0.045 & 0.026 \\ \hline
Quiescent Galaxies & $D1$ & $(0.5 < z < 1.0)$~~~to~~~$(1.0 < z < 1.5)$ & 0.050
& 0.17 \\
 & $D1$ & $(0.5 < z < 1.0)$~~~to~~~$(1.5 < z < 2.0)$ & 0.52 & 0.81 \\
 & $D1$ & $(1.0 < z < 1.5)$~~~to~~~$(1.5 < z < 2.0)$ & 0.46 & 0.61 \\ \hline
Star-Forming Galaxies & $D4$ & $(0.5 < z < 1.0)$~~~to~~~$(1.0 < z < 1.5)$ & 0.074 & 0.27 \\
 & $D4$ & $(0.5 < z < 1.0)$~~~to~~~$(1.5 < z < 2.0)$ & 0.23 & 0.82 \\ 
 & $D4$ & $(1.0 < z < 1.5)$~~~to~~~$(1.5 < z < 2.0)$ & 0.38 & 0.96 \\ \hline
Star-Forming Galaxies & $D1$ & $(0.5 < z < 1.0)$~~~to~~~$(1.0 < z < 1.5)$ & 0.027
& 0.30 \\
 & $D1$ & $(0.5 < z < 1.0)$~~~to~~~$(1.5 < z < 2.0)$ & 0.12 & 0.74 \\
 & $D1$ & $(1.0 < z < 1.5)$~~~to~~~$(1.5 < z < 2.0)$ & 0.20 & 0.59 \\ \hline
\enddata
%
\tablecomments{\editone{(1) Either quiescent or star-forming galaxies,
  using the definition in \S~\ref{section:data}.  (2) Environmental
  density quartile, where $D4$ is the highest density quartile and
  $D1$ is the lowest density quartile (see \S~\ref{section:env}).  (3)
  The redshift ranges of the galaxy subsamples compared in the
  tests. (4) The $p$--value from the KS test.  (5) The $p$--value from
  the MWW test.  The $p$--value is a likelihood that the samples are
  drawn from the same parent distribution. A low $p$-value implies
  greatly likelihood for differences in  the stellar--mass
  distributions between the galaxy subsamples.}}
\end{deluxetable*}

\editone{To support these conclusions, we have applied non-parametric
statistical tests to the distribution of stellar masses of the
galaxies in the different subsamples.   Specifically, we used the
Kolmogorov--Smirnov (KS) and Mann--Whitney--Wilcoxon  (MWW) tests
\citep[see][]{feig12} as implemented in R to test the
hypothesis that the (unbinned) distributions of stellar mass for the
different galaxy subsamples are drawn from the same parent
distribution.     In all cases we consider the distribution of stellar
masses for galaxies down to the stellar--mass completeness limit at
each redshift. We also consider only the evolution of the
stellar--mass distributions within ZFOURGE to minimize systematics (as
all quantities are derived internally from the same dataset) and
because it is at $z > 0.5$ within the ZFOURGE data that the impact of
environment is most apparent (see figures~\ref{fig:smf} and
\ref{fig:smf_ratio}).  Table~\ref{table:pvalues} lists results (the
$p$--values) of the  KS and MWW tests comparing the different galaxy
stellar--mass distributions.  Here we focus on the results for
quiescent galaxies.  Results for star-forming galaxies are described
in Appendix~\ref{section:appendix}.}

\editone{We first compared the unbinned stellar--mass distribution of
  quiescent galaxies in the highest environmental density quartile
  ($D4$) at $0.5 < z < 1.0$ to the stellar--mass distributions of
  quiescent galaxies in $D4$ at higher redshift, $1.0 < z < 1.5$ and
  $1.5 < z < 2.0$ (for completeness, we also compare the stellar--mass
  distributions  between the galaxies at $1.0 < z < 1.5$ and those at
  $1.5 < z < 2.0$, and in all cases find consistent results).   Comparing the
  stellar--mass distribution of the $0.5 < z < 1.0$ galaxies to that
  at $1.0 < z < 1.5$, the KS test gives $p = 6.6\times 10^{-5}$ and
  the MWW test gives $p = 8.3\times 10^{-4}$.   Comparing the
  stellar--mass distribution of the $0.5 < z < 1.0$ galaxies to that
  at $1.5 < z < 2.0$, the KS test gives $p = 1.0\times 10^{-5}$ and
  the MWW test gives $p = 4.6\times 10^{-6}$.  In all cases we would
  reject the hypothesis that these stellar mass distributions are
  drawn from the same parent distribution at $>$99.9\% significance.   This
  supports the claim that there is strong evolution in the shape of
  the distribution of stellar masses with redshift for quiescent
  galaxies in the highest environmental densities. } 

\editone{In contrast, comparing the unbinned stellar--mass
distribution of quiescent galaxies in the lowest density quartile
($D1$) at $0.5 < z < 1.0$ to the stellar mass distribution of these
galaxies $1.0 < z < 1.5$ the KS test gives $p = 0.05$ and
the MWW test gives $p = 0.17$.  This provides very weak
evidence that they are drawn from different parent distributions
(equivalent to $\approx 1.5\sigma$ assuming a Gaussian distribution),
but could imply some evolution in the shape of the SMF if confirmed by
larger datasets.  Comparing the
stellar--mass distribution at $0.5 < z < 1.0$ to the distribution at
$1.5 < z < 2.0$,  the KS test gives $p$=0.52 and the MWW test gives
$p$=0.81.    These tests support the conclusion that there is very
little measurable evidence for redshift evolution in the shape of the
distribution of stellar masses for quiescent galaxies in the lowest
environmental densities. }

To quantify the evolution in the  SMFs, we fit them with \citet{sche76}
functions, defined as
\begin{eqnarray}
\phi(m)\ dm =&\ln(10)\ \phi^\ast\ 10^{(m - m^\ast)(1+\alpha)}\
\\ \nonumber 
&\times \exp(-10^{(m-m^\ast)})\ dm,
\end{eqnarray}
where again for convenience we define $m = \log_{10} M_\ast/\msol$ as the
logarithm of the stellar mass, and $m^\ast=  \log_{10} M^\ast/\msol$ as the
logarithm of the characteristic stellar mass.  For the fitting, we
only included data where the SMFs are complete in stellar mass for
quiescent galaxies using the values from \citet{kawi17}.   We fit using a
Markov Chain Monte Carlo (MCMC) similar to \citet{fore13} and adapted
to IDL by R. Russell (2015, private communication). 

\begin{figure*}
\epsscale{0.9}
\plotone{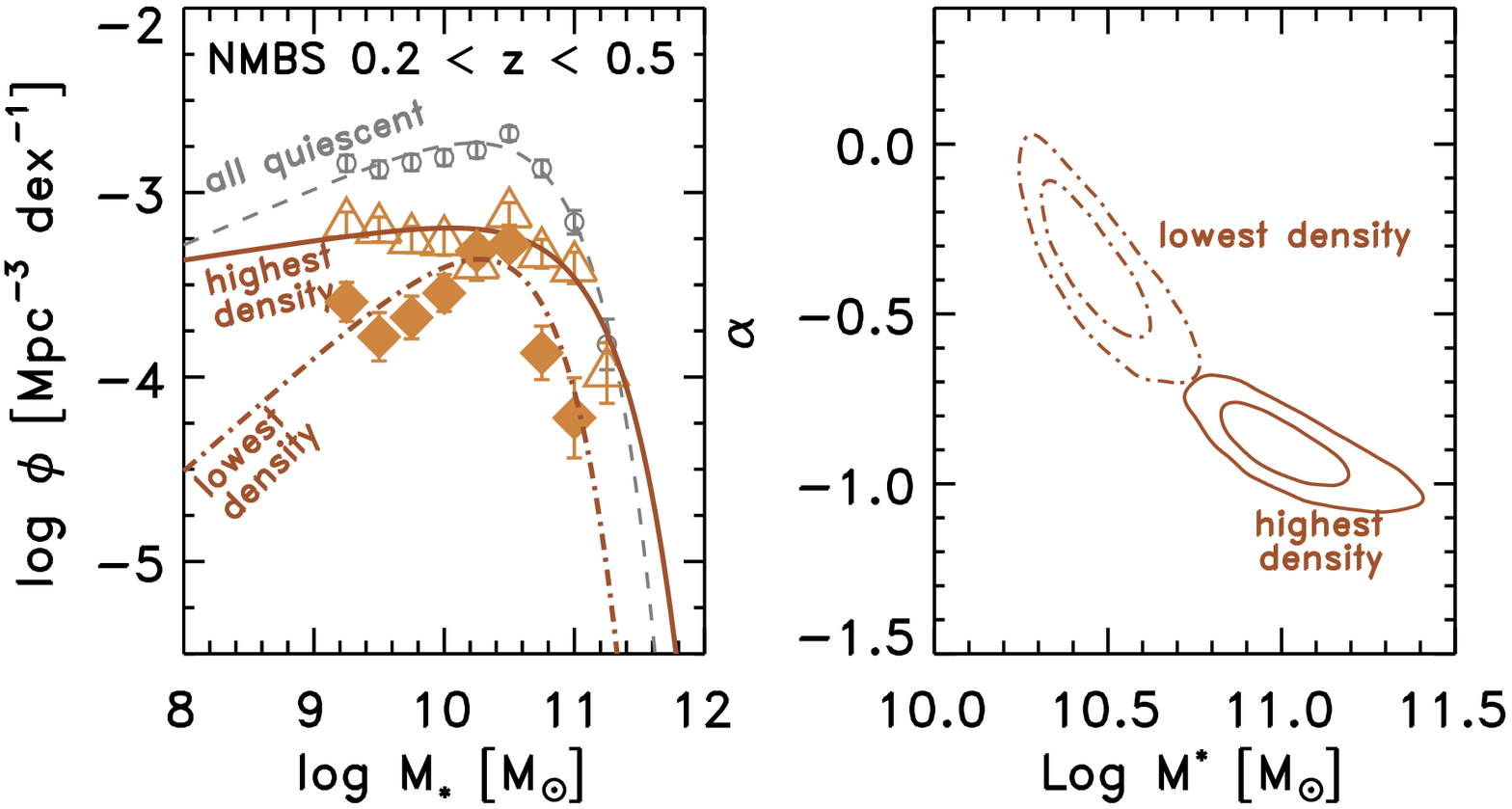}
\plotone{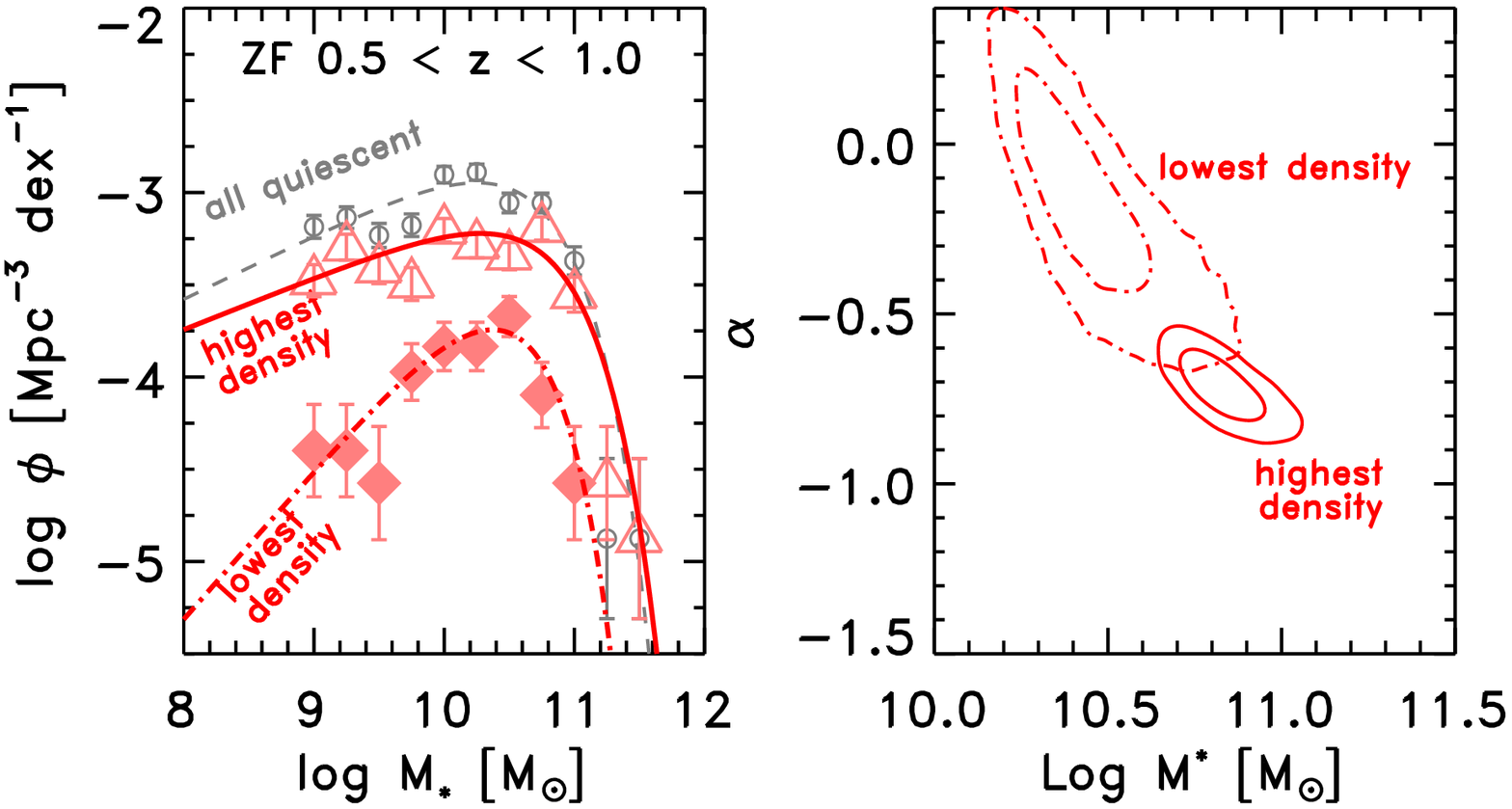}
\epsscale{1.0}
\caption{ \citet{sche76} model fits to the Stellar Mass Functions
  (SMFs) for quiescent galaxies.   The top--left panel shows the SMF
  for quiescent galaxies at $0.2 < z < 0.5$ in the NMBS.   The lines
  show the best-fitting Schechter model for the SMF for all quiescent
  galaxies, and for quiescent galaxies in the highest and lowest
  density quartiles, as labeled.    The top--right panel shows the
  68\% and 95\% confidence regions on the 
  characteristic stellar mass ($M^\ast$) and the low-mass slope
  ($\alpha$) for galaxies in the highest and lowest density
  quartiles.   The bottom panels show the same information for
  quiescent galaxies in the ZFOURGE (ZF) survey at $0.5 < z < 1.0$.  In
  both samples there are significant differences in the model fits for
  the SMF of  quiescent galaxies in the highest and  lowest
  densities, particularly in the low-mass slope.   \label{fig:smf_fit}}
\end{figure*}

Figure~\ref{fig:smf_fit} shows example results of the
Schechter--function fits to the SMFs for quiescent galaxies from NMBS
and ZFOURGE out to $z < 1$ (see below for discussion of results for
higher redshifts).   The panels show that out to $z < 1$ the SMF for
quiescent galaxies in the highest density regions ($D4$) has (1) a
higher characteristic mass ($\log M_\ast$), accompanied by a higher
normalization ($\phi^\ast$), and (2) a steeper low-mass slope
($\alpha$), relative to the quiescent galaxies in the lowest density
regions ($D1$).  These results are significant at $>$99\% confidence
(i.e., there is a $<$1\% likelihood that $\log M_\ast$ and $\alpha$
are identical).  The fact that the results are the same from the two
independent datasets, NMBS at $0.2 < z < 0.5$ and ZFOURGE at $0.5 < z
< 1$, increases this significance. This makes the prediction that
future surveys, covering more area to our depth will  reinforce this
conclusion.  

\begin{figure*}
\epsscale{0.9}
\plotone{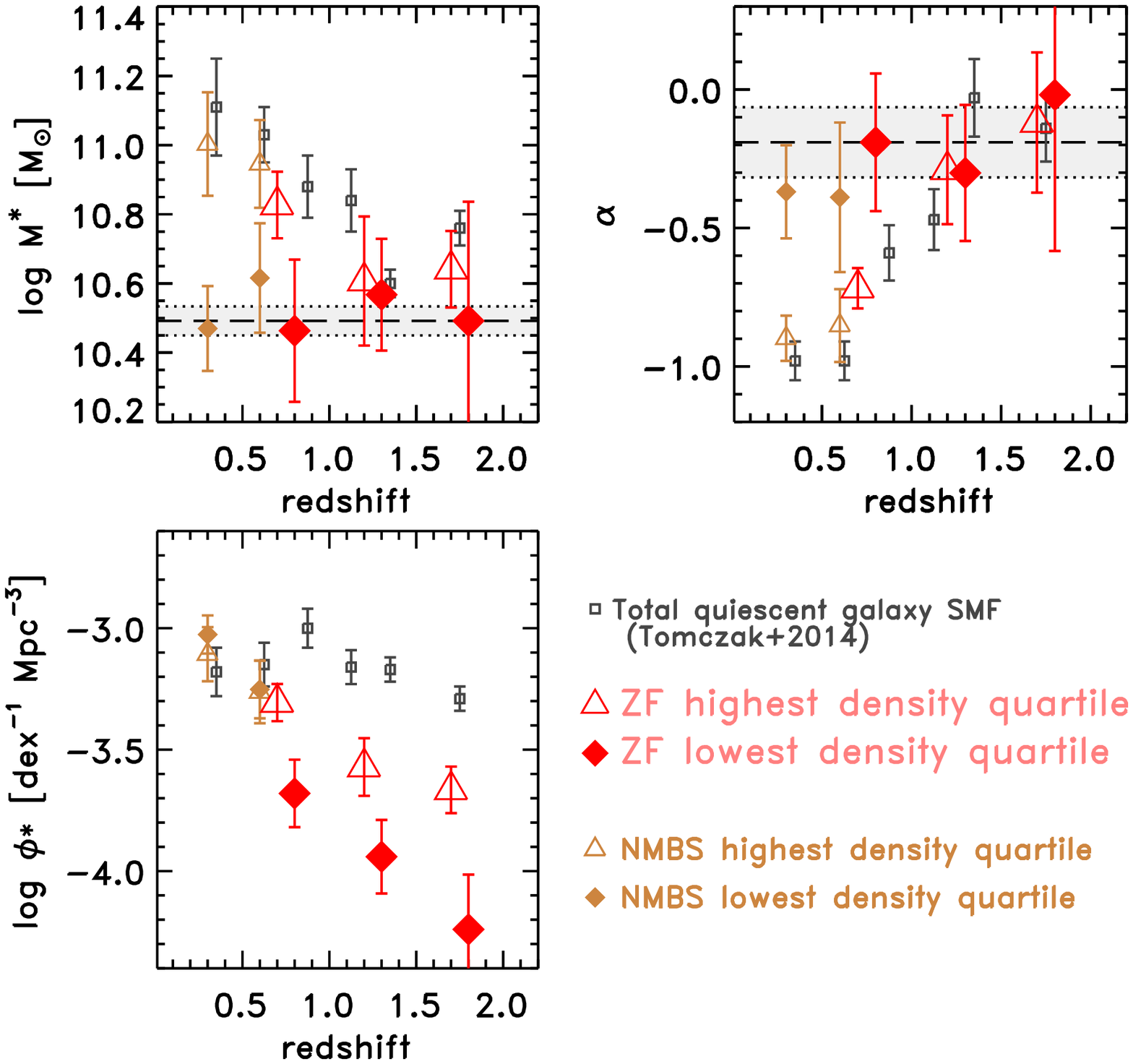}
\epsscale{1.00}
\caption{ Evolution of the  SMF model parameters for quiescent
galaxies as a function of redshift and galaxy density.   The panels
show the evolution with redshift for the characteristic stellar mass
($M^\ast$)), the low-mass slope ($\alpha$), and the normalization
($\phi^\ast$)  from the fits to SMF of quiescent galaxies in NMBS and
the ZFOURGE survey.    The symbols correspond to the same SMFs in each
panel, as labeled.  For comparison, the small open, black squares show
the evolution of model parameters for the total SMF of quiescent
galaxies measured by \citet{tomc14}.  In the top panels, the
gray-shaded region shows the mean range of mean values for $M^\ast$
and $\alpha$ for the low-density environments, which is consistent
with no evolution in redshift.      The quiescent galaxies in the
highest density quartile track the evolution in the total quiescent
galaxy SMF.  By contrast there is no evidence for \textit{any}
evolution \editone{in the shape of the SMF} for quiescent galaxies in
the lowest--density quartile (defined by $\log M^\ast$ and $\alpha$),
down to the ZFOURGE stellar mass limit, other than an \editone{overall
increase in number density} ($\phi^\ast$).     \label{fig:smf_parm}}
\end{figure*}

Figure~\ref{fig:smf_parm} shows the redshift evolution of the fitted
parameters ($\phi^\star$, $M^\ast$, $\alpha$) of the SMFs for the
quiescent galaxies in the highest and lowest density quartiles.   We
noted above that the shape of the SMF of quiescent galaxies in the
lowest density quartile is consistent with no evolution.
Figure~\ref{fig:smf_parm} shows that this is quantitatively true: the
quiescent galaxies in the lowest density regions show no evidence of
evolution in $M^\ast$ or $\alpha$ over the entire redshift range, $0.2
< z < 2.0$, nor is there (statistically) any indication that the
parameters change between the independently processed NMBS and ZFOURGE
datasets.  (In Appendix~\ref{section:appendix} we show there is no
evidence for evolution in $M^\ast$ or $\alpha$ for star-forming
galaxies.) 

In fact,  the evolution in the shape of the of the SMF of quiescent
galaxies in the highest density environments tracks that of the total
quiescent--galaxy SMF.   \citet{tomc14} showed that the shape of the
quiescent galaxy SMF evolves over this redshift range, and here we
show this is due to evolution in overdense regions, presumably from
evolution in environmental quenching.   Figure~\ref{fig:smf_parm}
shows that $M^\ast$ and $\alpha$ evolve strongly for the quiescent
galaxies in the highest density regions ($D4$), and these match the
observed evolution in the \textit{total} quiescent galaxy SMF measured
independently \citep[though with an earlier version of the ZFOURGE
dataset,][]{tomc14}.  The figure also shows that there is an overall
increase in the normalization ($\phi^\ast$) of the SMF in all
environments, as expected as structure grows in all cosmic densities
\citep[\eg,][]{spri05a}.


To summarize these findings, (1) there is no evidence that the SMF of
star-forming galaxies depends strongly on environment at any redshift
(with some difference in the number density of massive star-forming
galaxies);  (2) the SMF of quiescent galaxies \emph{does} depend strongly on
environment; \editone{(3), while there is no evidence that the shape of the SMF of quiescent
galaxies in the lowest density regions ($D1$) from $z=2.5$ to 0.2, we 
do see strong evolution in the highest density regions ($D4$). This latter point
suggests that the observed evolution in the SMF for the full population of
quiescent galaxies is limited to densities where environmental effects are prominent.}

\section{Discussion}\label{section:discussion}

\subsection{The Dependence of the Stellar Mass Function on Environment}

One of our main conclusions is the rapid increase in the number
density of lower mass ($\log M_\ast/\msol \simeq 9-10$) quiescent
galaxies in denser environments over the redshift range from $z\simeq
1.5$ to $z\simeq 0.2$.  Previous studies have made similar claims
\citep[e.g.,][]{bolz10,vulc12,quadri12,vanderburg13,davi16,nant16,ethe17}.
Many of these studies have been restricted by their stellar mass
limits to $\log M/\msol \gsim 10$, where our results are complete to
stellar masses to $\log M/\msol \simeq 9-9.5$.  This provides better
constraints in the SMF fitting (particularly for the $M^\ast$ and
$\alpha$ parameters) as the ZFOURGE data are complete to more than
$\approx$1 dex below the characteristic stellar mass, $M^\ast$.  Our
results are consistent with the suggestion from \citet{mort15}, who
found tentative evidence for a higher number density of quiescent low
mass galaxies in denser environments in this redshift range.

\citet{balogh16} reported a possible lower number density of lower
mass quiescent galaxies in groups at $0.8 < z < 1.0$, while the SMF of
quiescent galaxies in rich clusters at $z\sim 1$ showed no difference
to the SMF in the field \citep[similar to the findings of][using the
same data]{vanderburg13}.    However, both studies (and others)
compared their SMF of cluster galaxies to  ``field'' galaxies from the
COSMOS/UltraVISTA data covering  1.6 deg$^2$ \citep{muzzin13c}.  As
discussed in \S{\ref{section:intro}}, this poses a complication as
UltraVISTA includes many overdense regions similar to the ones in the
higher-density quartiles in our analysis.  One of our findings is that
quiescent galaxies in highest-density regions in such (``field'')
survey data dominate the shape of the \textit{total} quiescent galaxy
SMF, and \editone{this could explain} the lack of difference between the
cluster and field SMF, and may compromise studies that compare
clusters/groups to ``field'' data that includes such structures.
Furthermore, it suggests that the dominant environmental effects
driving the evolution of the quiescent galaxy SMF are apparent in
overdensities similar to group environments, before the galaxies are
accreted into clusters \citep[see also][]{mcgee09,foss17}.
Therefore, we predict that if one restricts the analysis of ``field''
surveys (such as COSMOS/UltraVISTA) to only regions of lower-than-mean
density, one would find results consistent with those we report here. 

At the high stellar-mass end, there is also evidence for evolution in
the shape of the quiescent galaxy SMF with environment, particularly
at $z < 1$.  Figure~\ref{fig:smf_parm} shows this as the evolution of
the characteristic mass, $M^\ast$, from the parametric fits to the
SMFs.  Beginning at the highest redshifts, $1.5 < z < 2.0$, the
characteristic masses of the SMF in the highest and lowest density
quartiles are consistent with each other.  Differences develop over
time (decreasing redshift), where at $z < 1.5$ we find that $M^\ast$
is larger in the highest density regions compared to the lowest
density regions, reaching a difference of $\sim$0.5 dex (factor of
order 3) by $z \sim 0.5$.  The fact that this is observed in our
analyses of the independent NMBS and ZFOURGE datasets reinforces the
significance of this result \citep[this is also consistent with the
  findings of][]{mort15}.

Remarkably, in the lowest--density regions, there is no indication
that the \textit{shape }of the SMF evolves for quiescent galaxies.
This is seen in Figure~\ref{fig:smf_parm} as a lack of observed evolution in
$M^\ast$ or $\alpha$ for quiescent galaxies in low densities.  This
means that any mass growth for quiescent galaxies appears to be
isolated to higher density regions.     The lowest density
environments continue to produce quiescent galaxies, as we do observe
an increase in $\phi^\star$ with decreasing redshift.  This can be
explained by a universal process that continuously quenches galaxies,
producing new quiescent galaxies, and acts in all density regions
\citep[\eg, this may be related to halo quenching of all
  galaxies above a mass threshold,][]{dekel06a,catt08}.   

Much of the growth in quiescent galaxies is expected to occur through
non-dissipative (``dry'') mergers \citep[\eg,][]{oser10,vandokkum10}.
Our findings therefore suggest that these mergers occur only rarely
for galaxies in low density regions (or we would expect evolution in
$M^\ast$).  In contrast, we do observe strong $M^\ast$ evolution 
for quiescent galaxies in higher density regions, starting at $z\gsim
1$ (Figure~\ref{fig:smf_parm}).   This is consistent with studies that
advocate for enhanced or accelerated growth of quiescent galaxies in
high-density regions at high redshifts
\citep[\eg,][]{papo12,rudn12,andr13,bass13,newman14}. 
%
%
Furthermore, the differences in $M^\ast$ for quiescent galaxies as a
function of environment appear at the epoch when we expect rapid
collapse of groups and clusters \citep[e.g., $z \sim 1-1.5$,
see][]{muld15}, and we may be witnessing enhanced mergers as galaxies
in these structures coalesce \citep[e.g,][]{lotz13}.   At lower
redshifts $z < 1$, this trend continues, and quiescent galaxies in
higher density regions continue to grow, likely through mergers
between satellites and centrals \citep[as has been measured in some
studies of cluster/group galaxies,
e.g.,][]{tran05,bundy09,lidman12,tomc17}, or through major mergers of
more massive star-forming galaxies that also occur more frequently in
denser environments \cite[\eg,][]{hopk10a,davies16}. These merger
events may be nearly non-existent for quiescent galaxies in
low-density environments. 

For star-forming galaxies, there is no measurable difference in the
evolution of the SMF with environment over the range $0.2 < z < 2.0$
(e.g., Figure~\ref{fig:smf}).    This is consistent with previous
findings from $0.02 < z < 1.3$ that the shape of the SMF for
star-forming galaxies is mostly independent of environment
\citep{peng10,giod12,vanderburg13,davi16}.  Other studies at redshifts
$z \gsim 0.2$ counter this with evidence that the SMF of star-forming
galaxies shows a higher normalization or shift to higher
characteristic stellar masses in denser environments (X-ray groups and
clusters) compared to the field \citep{giod12,mok13,davi16,tomc17}.   A
reason for these differences may be  differences in the definition of
environment, as here we use the relative overdensity rather than
quantities that scale with halo mass (such as X-ray luminosity or
velocity dispersion) as used in the other studies. To rectify this
possible difference will require a critical comparison of the results
from both methods applied to the same datasets, which is beyond the
scope of this work.   Regardless, in our study environment appears to
play at most a very small role in shaping the SMF of star-forming
galaxies, except perhaps in the highest densities of richer clusters
than we probe here.

\subsection{Implications for the Environmental Quenching Efficiency}

One contentious point in the literature is whether the effects of mass
quenching (quenching associated with processes internal to the galaxy
that scale with stellar mass) and environmental quenching  (quenching
associated with changes in the average galaxy density) are separable
(i.e., non--covariant;  \S~\ref{section:intro}).  If they are, then it
implies that the same environmental effect(s) act on all galaxies
independent of their stellar mass, and this in turn constrains the
dominant environmental process(es).    While there is strong
observational evidence that the effects are separable at low-redshifts
\citep[e.g.,][]{peng10}, there is growing evidence that the
environmental quenching includes a dependence on galaxy stellar mass
at high redshifts \citep{balogh16,kawi17}.

Previous studies quantify the mass quenching  and environmental
quenching in terms of the quenching \textit{efficiency}.  The
mass-quenching efficiency is then defined as the excess quenching with
increasing stellar mass, holding environment (overdensity) fixed;  the
environmental--quenching efficiency is the excess quenching with
increasing galaxy overdensity, holding stellar mass fixed
\citep[\eg,][and other references in \S~1]{peng10}.\footnote{Various
naming conventions for what we call the ``environmental quenching
efficiency'' exist in the   literature, including ``transition
function'' \citep{vandenbosch08} and ``conversion fraction''
\citep{phil14,foss17}. }  

Using ZFOURGE data, \citet{kawi17} measured an environmental quenching
efficiency from the fraction of quenched galaxies as a function
stellar mass and overdensity.   They showed that at higher redshift,
$z \gsim 0.5$, the environmental quenching efficiency must decline
with decreasing stellar mass to account for the low fraction of lower
mass quiescent galaxies.  Our results for the evolution of the galaxy
SMF requires an environmental quenching efficiency that both (1)
evolves with time and (2) depends on stellar mass.  If this were not
the case (and the environmental quenching efficiency was constant with
stellar mass), then the shape of the quiescent galaxy SMF in high
density environments would appear quite different from the one
observed.  The reason for this is that the low-mass slope of the
\textit{star-forming} galaxy SMF is very steep, $\alpha \sim -1.2$ to
$-1.5$, identical in high density and low density regions
(Figure~\ref{fig:smf}),  and nearly unchanging with redshift
\citep[Figure~\ref{fig:smf}; see also][]{tomc14}.    If the
environmental quenching efficiency were constant (mass-invariant),
then the low-mass end of the \textit{quiescent} galaxy SMF should show
a similarly (steep) low-mass slope, which is not supported by the
data.  

\begin{figure*}
\epsscale{1.2}
\plotone{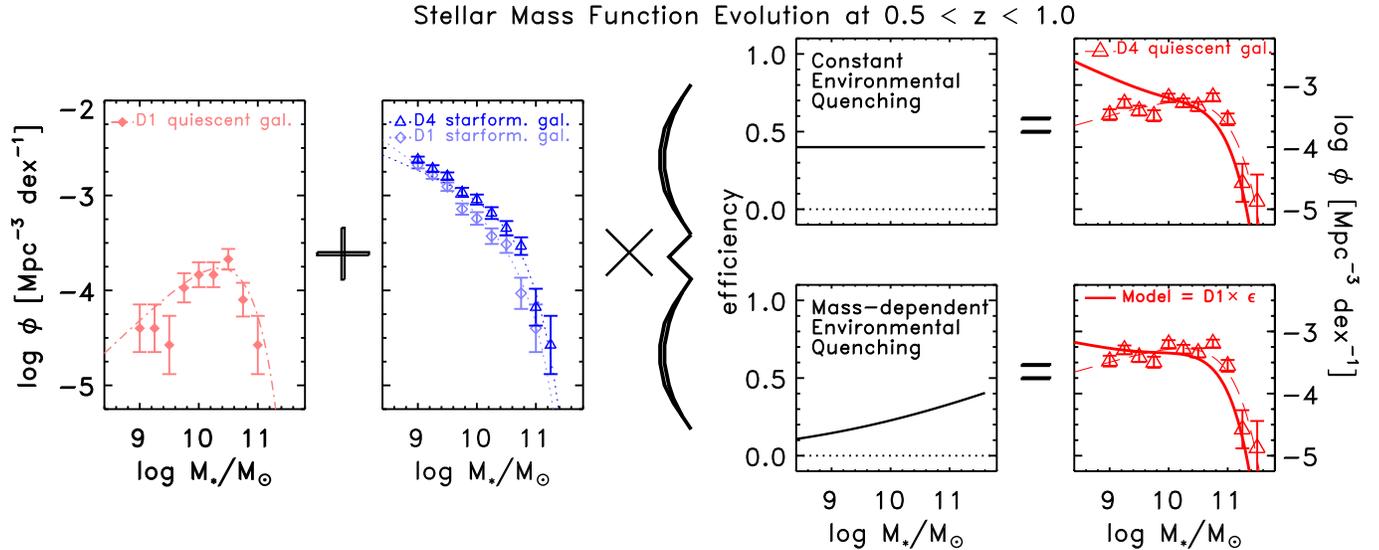}
\epsscale{1.00}
\caption{Simple experiments of the evolution of the stellar-mass function (SMF) of quiescent
galaxies at $0.5 < z < 1.0$ in different environments.   The models
demonstrate that the environmental quenching must be dependent on the
stellar mass to reproduce the SMF of quiescent galaxies in different
environments.  \editone{ The panels demonstrate an ``equation'', with
  some fraction of star-forming galaxies being quenched by the
  environment and added to the quiescent--galaxy SMF in the lowest
  density quartile ($D1$) to represent the quiescent--galaxy SMF in the
  highest density quartile ($D4$), as described in
  Eq.~\ref{eqn:smf_model} and \ref{eqn:smf_model2}.  }    The left panels show
the measured SMF  and Schechter-model fit for
quiescent and star-forming galaxies in the lowest overdensity quartile
($D1$) and the highest overdensity quartile ($D4$), as labeled.  
The right-hand panels show results derived
by adding quenched star-forming galaxies using different environmental quenching
efficiencies.  The lower row of panels show an environmental quenching
efficiency that depends on stellar mass from \citet{kawi17}, and the
upper panels show results derived using an environmental quenching efficiency that is
constant with stellar mass \citep[with values from][]{peng10,kovac14}.
A mass-independent environmental quenching efficiency would greatly
overproduce the number density of low-mass, quiescent galaxies in high
overdensities at $z > 0.5$.  The environmental quenching efficiency must
decline with decreasing stellar mass at these redshifts.  \label{fig:smf_model}}
\end{figure*}

\begin{figure*}
\epsscale{1.2}
\plotone{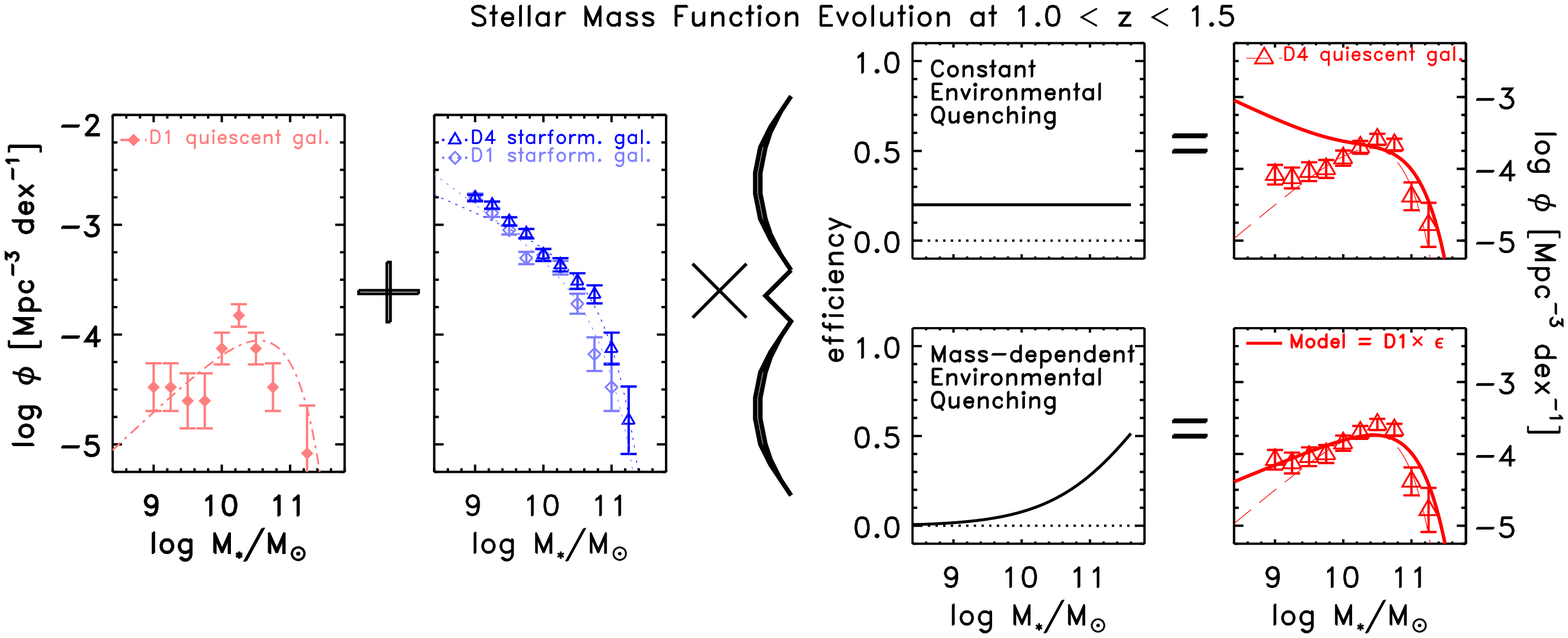}
\epsscale{1.00}
\caption{\editone{Same as Figure~\ref{fig:smf_model} but for galaxies at $1.0
  < z < 1.5$.    Because the low-mass end of the star-forming galaxy
  SMF is so steep, an environmental quenching efficiency that is
  constant in stellar mass would greatly overproduce the number
  density of low-mass, quiescent galaxies in high overdensities, even
  more dramatically at this redshift than at $0.5 < z < 1.0$. \label{fig:smf_model2}}}
\end{figure*}
 
\editone{Figures~\ref{fig:smf_model} and \ref{fig:smf_model2} illustrate this point} with a simple
experiment.   At both $0.5 < z < 1.0$ and $1.0 < z < 1.5$ we model the
quiescent galaxy SMF in the highest density regions ($D4$) by summing
the quiescent galaxy SMF in the lowest density regions ($D1$) with a
fraction of the  SMF of star-forming galaxies.  This simple model
represents the sum of those galaxies that have been quenched by mass
only (argued to be the case for $D1$) with the fraction of
star-forming galaxies recently quenched by their environment,
represented by the environmental quenching efficiency. Note that this
experiment excludes other effects, such as merging between galaxies,
which could change the relative number of low-mass and high-mass
galaxies \citep[see discussion above and, e.g.,][]{tomc17}. 
Specifically, for the case of a mass--constant environmental quenching
efficiency ($d\epsilon/dM_{\ast}$=0) we take, 
\begin{equation}\label{eqn:smf_model}
\phi(M_\ast)_{Q,D4} = \phi(M_\ast)_{Q,D1} + \phi(M_\ast)_{\mathrm{SF}}
\times \epsilon_\mathrm{const}, 
\end{equation}
where $\phi(M_\ast)_{Q,D4}$ is the modeled quiescent galaxy SMF in the
highest density regions, $\phi(M_\ast)_{Q,D1}$ is the measured
quiescent galaxy SMF in the lowest density regions,
$\phi(M_\ast)_{\mathrm{SF}}$ is the star-forming galaxy SMF (in either
the highest or lowest density regions as they are so similar), and
$\epsilon = \epsilon_\mathrm{const}$ is a
mass--invariant environmental quenching
efficiency. Figure~\ref{fig:smf_model} shows this substantially
overproduces the number density of low-mass quiescent galaxies in the
highest density regions derived from the data \editone{at $0.5 < z <
  1.0$.  Figure~\ref{fig:smf_model2} shows this effect is more
  pronounced at $1.0 < z < 1.5$ where the low-mass end of the
  quiescent galaxy SMF is even shallower and quenching of star-forming
  lower-mass galaxies would have even more impact.}

In contrast, \editone{Figures~\ref{fig:smf_model} and
\ref{fig:smf_model2} show} that an environmental quenching efficiency
that increases with stellar mass reproduces the quiescent galaxy SMF
in the highest density regions.   For this calculation, we use
\begin{equation}\label{eqn:smf_model2}
\phi(M_\ast)_{Q,D4} = \phi(M_\ast)_{Q,D1} + \phi(M_\ast)_{\mathrm{SF}}
\times \epsilon(M_\ast),
\end{equation}
where all variables are the same as in Equation~\ref{eqn:smf_model}
except the environmental quenching efficiency $\epsilon(M_\ast$) now
varies with stellar mass, where we have taken the measurements from
\citet{kawi17}.   Figure~\ref{fig:smf_model} shows that this
environmental quenching efficiency that decreases with decreasing
stellar mass qualitatively reproduces the quiescent galaxy SMF in the
highest density regions.\footnote{This is partly by construction as we
have used the same ZFOURGE dataset to derive the SMFs and the
quenching efficiencies in \citet{kawi17}, but the point remains that
the environmental quenching efficiency must decrease with decreasing
stellar mass at high redshift.}    The implication is that at $0.5 < z
< 1.0$ and $1.0 < z < 1.5$ the environmental quenching efficiency
\textit{must} depend on the stellar mass of the galaxies.

There are reasons that our result is seemingly at odds with some
previous studies.  At $z < 0.2$ the measured environmental quenching
efficiency is nearly constant with stellar mass \citep{peng10,whee14}.
At higher redshift, $0.4 < z < 0.7$, some studies argued for a
similarly (mass-invariant) constant environmental quenching efficiency
\citep{peng10,kovac14}.  However, these conclusions are limited to
relatively moderate stellar masses, $\log M_\ast/\msol > 10.3$ at
$z=0.7$, where our analysis shows that a constant environmental
quenching efficiency is able to reproduce observed evolution in the
SMF  (see Figure~\ref{fig:smf_model}).    It is only by probing to
lower stellar masses (where with ZFOURGE we are complete to $\log
M_\ast/\msol \simeq 9$ at $z=2$:  nearly 1 dex below the
characteristic stellar mass, $M^\ast$)  that the mass--dependence of
the environmental quenching efficiency becomes apparent.  Indeed,
\citet{kovac14} acknowledge this possibility, where they stated that
they ``cannot exclude the existence of a cross term in mass and
environment, but this must be within [the] uncertainties''.  Our
results show evidence for such a cross term exists at lower stellar
masses ($\log M_\ast/\msol \lsim 10.3$).

The environmental quenching efficiency likely also  increases in richer 
environments.   This is true at low redshifts \citep[$z <
0.2$,][]{peng10,whee14}, and at high redshift ($z\sim 1-1.5$) where
\citet{balogh16} and \citet{nant17} reported evidence that the
environmental efficiency in clusters is higher than in groups or in
the field.    As discussed above (\S~\ref{section:env}), the
environments (halos) of objects in our high density regions likely
correspond to groups (with few, if any ``clusters'').  This suggests that
the physical processes that produce the high environmental quenching
efficiency act in such group--sized environments \citep[see also][]{foss17}.  At
high redshifts this means the environmental quenching
efficiency depends on both stellar mass \textit{and} environment
(i.e., the halo mass).  For
example, at fixed stellar mass, e.g., $\log(M_\ast/\msol) \simeq
10.5$, the environmental quenching efficiency that we require for the
evolution of the SMF, $\epsilon \simeq 0.2-0.3$, is
consistent with that found in groups by \citeauthor{balogh16}.   At
lower stellar masses, \citeauthor{balogh16} find tentative evidence
that the environmental quenching efficiency declines, similar to what
we require to account for the evolution of the SMF. 

Alternatively, one must consider that the evolution we attribute to a
mass--dependent environmental quenching efficiency is somehow
connected to our use of photometric redshifts.    We argue that this
is not the case.  Our analysis of the relative and absolute
uncertainties in the ZFOURGE photometric redshifts show they are are
low ($\sigma_z / (1+z) \sim 0.01-0.02$)  for galaxies to our stellar
mass limit \citep{kawi14,tomc14}, with no significant differences for
quiescent and star-forming galaxies \citep{stra16}.  These
uncertainties are consistent with comparisons to spectroscopic
redshifts \citep[for emission-line sources at $1 \lsim z \lsim
  2$,][]{nana16}. This redshift accuracy is
sufficient to identify regions of high and low density
\citep[\eg,][]{spit12,mala16}.
%
%
This is consistent with the analysis of \citet{kawi17}   who showed
that with such precise photometric redshifts, one recovers accurately
galaxies in the highest and lowest density quartiles compared to
densities measured with typical errors of spectroscopic redshifts.
Therefore, we do not expect the use of the photometric redshifts to be
a dominant driver of our results, but this must be tested by large
spectroscopic datasets that include redshifts for quiescent
(absorption--line) galaxies down to the magnitude limit of our stellar
mass limit \citep[$\ks \sim 25$~mag; see][]{stra16}.  

\subsection{Implications for the Environment Quenching Process(es)}

Our results show the evolution of the quiescent galaxy SMF requires
an environmental quenching efficiency that depends on stellar mass at
redshifts $z > 0.5$.   In contrast, results at $z < 0.2$ show that the
correlations between stellar mass and environment on quenching are
separable  \citep{peng10,whee14}.   This suggests that the efficiency
of at least one of the physical processes responsible for
environmental quenching changes with time. At present, it might then
be a cosmic coincidence that the environmental quenching efficiency
appears mass invariant, or it could indicate that environmental
quenching at present is dominated by a process that is mostly
independent of stellar mass.    We consider physical processes that
would explain these observations.    

Models for environmental quenching fall broadly into two classes,
those that depend on the mass of the halo (and are related to the
dynamical time of the satellite galaxy and the halo mass of the
central) and those that depend on the properties of the satellite galaxy
(and are related to the satellite's stellar or halo mass).  In
addition, reality may require a combination of the two that depend on both the
stellar and halo masses of the central and satellite.  

Environmental physical processes that depend on the mass of the halo
include gas stripping of both the gaseous halo and/or the interstellar
medium of the satellite
\citep[e.g.,][]{gunn72,dekel06a,mccarthy08,tonn09}, or tidal
interactions and galaxy mergers
\citep[\eg,][]{faro81,dekel03,deason14}.  The magnitude of these
environmental processes depends on the mass of the central galaxy's
halo, and the amount of time a galaxy spends as a satellite, with a
weaker dependence on the satellite's stellar mass.  It is difficult to
estimate the impact of the environmental quenching mechanisms
theoretically.  For example, models generally have difficulty
reproducing the star-formation distribution of observed satellites
\citep{font08,wein10,mccarthy11}, but they can account for quenching
in lower-mass satellites that are more susceptible
to effects such as gas stripping and strangulation
\citep[][]{wetzel15}.  Dynamical friction may also cause massive
galaxies to segregate toward the center of the potential well of a
group, where merging episodes are more likely to occur \citep[see
  discussion in][]{giod12}.  This may explain the strong redshift
evolution in the characteristic mass of quiescent galaxies in the
highest density regions (Figure~\ref{fig:smf_parm}).

Indirectly, the environmental processes such as strangulation and
starvation, where a galaxy's supply of cold gas is cut off, and/or the
hot gaseous halo is removed as the galaxy becomes a satellite
\citep{larson80,balogh00,feld11}, can lead to a dependence on the
stellar mass of the satellite.  In a process dubbed
``overconsumption'', \citet{mcgee14} demonstrated that the combination
of star formation and star-formation-driven outflows in satellites
leads to shorter gas depletion timescales.  A satellite then quenches rapidly if the halo of the
central prevents the accretion of additional gas \citep{dekel06a}.
This process of overconsumption is more efficient at high redshifts
where galaxy star-formation rates (SFRs) are generally higher, and it
leads to faster quenching in
more massive star-forming galaxies \citep{noeske07a,tomc16}, because
they have shorter gas--depletion timescales
\citep[\eg,][]{genzel15,papo16b,tacc17}.  \citeauthor{mcgee14} show
this leads to a predicted environmental quenching efficiency that
rises with stellar mass for star-forming satellite galaxies.   Our
results support a model like overconsumption as they require such a
relation between stellar mass and environmental quenching.

%
 \citet{balogh16} argue overconsumption produces the distribution of quenching and
consumption times as a function of satellite mass in both groups and
clusters at high redshifts.  Our results require that the
effects of overconsumption affect galaxies in 
(poorer) groups (not only rich clusters), as even the galaxies in our highest density
regions are most likely associated with such lower--mass structures \citep[see][and
discussion above]{foss17}.  Furthermore, overconsumption can easily act in
group environments at $z > 1$, as it requires only that halo of the central be of
sufficient mass to prevent gas accretion \citep[$M_h \sim
10^{12}$~\msol,][]{dekel06a}.   And most massive structures at $z > 1$
are galaxy groups, as clusters are still mostly in the act of
collapsing \citep{muld15,nant17}.  In their analysis of groups at $0.5 < z <
3.0$, \citet{foss17} advocate for environmental processes where
satellites exhaust their gas combined with an absence of gas accretion
\citep[akin to the ``starvation'' and ``overconsumption''
of][]{mcgee14}.  Even for clusters at $z\sim 1.5$, \citet{nant16} show
that the environmental quenching efficiency of galaxies has little
dependence on the cluster-centric distance out to 4 (projected) Mpc
\citep[see also][]{bass13}.  This further argues for processing of
galaxies in groups or cluster outskirts at these redshifts, which is
consistent with the interpretation of our results here. 

We conclude that a model such as overconsumption has the requisites to
be an effective quenching process for galaxies with stellar masses,
$\log M_\ast/M_\odot > 9$, at redshifts, $z > 0.5$.  At $z < 0.5$, the
fact that there is zero (or only a small) ``cross term'' between
environmental quenching efficiency and mass quenching efficiency
\citep[\eg][]{peng10,kovac14} compared to our results at higher
redshift argues that the strength of environmental processes are also
evolving with time.   This is likely through a combination of effects.
The strength of ``overconsumption'' is expected to be weaker at low
redshifts due to two factors \citep[see also][]{mcgee14}.    First,
satellite SFRs are lower \citep{madau14}, and gas--depletion times
longer \citep{genzel15}, which lengthens the timescale for galaxy
starvation.  Second, the dynamical time of the halos of central
galaxies become shorter relative to the Hubble time (and shorter than
the gas--depletion timescales).   This allows at late times for
alternative environmental processes to act that are associated with
the dynamical time of the central--galaxy halo (such as ram pressure
stripping), and the time a galaxy spends as a satellite \citep[and
indeed, there is evidence for longer quenching timescales for
satellites at $z\sim 0.5$ compared to satellites at $z\sim 1$ and
longer still at $z\sim 0$,][and see also, e.g., \citeauthor{tinker10}
\citeyear{tinker10}]{guo17}.  This evolution in the timescales
associated with gas--depletion and dynamical time explains both the
qualitative redshift evolution and dependence on the stellar mass for
the environmental quenching efficiency that we require to explain the
evolution of the SMF. 

Other environmental processes themselves may also evolve with time.  
For example, the model proposed by \citet{davies16} for galaxies at $z
< 0.1$ include different dominant environmental quenching processes
that depend on the satellite stellar mass. Interactions and mass
quenching dominate in the most massive galaxies ($\log M_\ast/M_\odot
> 10$), starvation dominates in moderate mass galaxies ($\log
M_\ast/M_\odot \simeq 8-10$), and ram-pressure stripping dominates in
lower mass galaxies \citep[$\log M_\ast / M_\odot < 8$][]{fill16}.  Our results point
to redshift evolution to this model, where the effects of starvation
(combined with shorter gas-depletion timescales) are more efficient in
quenching galaxies at higher redshifts, leading to an environmental
quenching efficiency that increases with stellar mass.  

%
%
%
%
%

\section{Summary}

We studied the evolution of the galaxy SMF over $0.2 < z < 2.0$ as a
function of galaxy activity -- quiescent compared to star-forming
galaxies -- and environment, using density measures derived from a
Bayesian-motivated distance to the third--nearest neighbor using data
from ZFOURGE and NMBS.    Our results extend to lower masses ($\log
M_\ast/M_\odot > 9.2$ [9.5] at $z=1.5$ [2.0]) than previously
possible.   The main results of our study can be summarized as
follows. 


%
For star-forming galaxies, there is no evidence the galaxy SMF depends
strongly on environment over the redshift range $0.2 < z < 2.0$ for
the stellar mass range of galaxies in our study.   The exception is
that there is some evidence for an excess number density of massive
($\log M_\ast/\msol \simeq 10.5$) star-forming galaxies in higher
density regions.   This may indicate the propensity of more massive
star-forming galaxies to be found in richer environments as has been
reported in some previous studies. 

%
For quiescent galaxies in the lowest density environments, the shape
of the SMF of quiescent galaxies shows no evidence of evolution over the
redshift range $0.2 < z < 2.0$ to our mass completeness limit,  other
than an overall increase in number density ($\phi^\ast$) with time.
This means that the mass-dependence of the non-environmental quenching
processes (i.e.\ mass quenching) does not evolve strongly with
redshift, and that arguably mass-quenching is the only mechanism for
producing quiescent galaxies of mass $\log M_\ast/\msol \gsim 9$ in
low-density regions out to $z < 2$.  

%
For quiescent galaxies in the highest density environments, the SMF at
$z \gsim 1.5$ is indistinguishable from that in the lowest densities.
Differences grow with decreasing redshift, and at $z \lsim 1$ the SMF
for quiescent galaxies in the highest density quartile shows higher
number densities compared to the SMF in the lowest density quartile,
particularly at low masses ($\log M_\ast/M_\odot \simeq
9-10$). Moreover, the evolution in the shape of the quiescent galaxy
SMF in the highest density environments (defined by $M^\ast$ and
$\alpha$ of the Schechter function) closely tracks that of the total
quiescent galaxy SMF.   Because the $M^\ast$ and $\alpha$ for the
quiescent galaxies in the lowest density environments --- where
environmental processes are expected to be minimal --- shows no
apparent evolution, we argue that environmental processes
are responsible for the evolution in the shape of the \emph{total}
quiescent galaxy SMF.

%
This evolution in the quiescent galaxy SMF
requires that the environmental quenching efficiency that depends on
stellar mass, such that the environmental quenching efficiency
decreases with decreasing stellar mass for $0.5 < z < 1.5$.   We show
with a simple model that if this were not the case (and the
environmental quenching efficiency were constant with stellar mass)
then it would overproduce the number of quiescent low-mass galaxies in
denser environments.   

%
We conclude that environmental processes that depend on galaxy
stellar mass (such as ``overconsumption''), where galaxies quench as
\editone{they become satellites} through a combination of rapid gas-consumption
and gas-ejection
timescales combined with the cessation of gas accretion from the
intergalactic medium, dominate environmental quenching in galaxies
with $\log M_\ast/M_\odot > 9$ at redshifts $z > 0.5$.    The fact
that the environmental quenching efficiency shows no dependence on
stellar mass at $z < 0.5$ argues that the relative strengths of
environmental processes evolve with time, and this is required to
account for the observed evolution in the galaxy SMF.  A physical
explanation for the evolution of the environmental processes is that at fixed
stellar mass, satellite SFRs decrease with decreasing redshift (and
gas-depletion times increase), making processes such as
overconsumption less efficient.  At the same time, the dynamical times
of satellites in the halos of central galaxies become shorter relative
to the Hubble time, which allows more time for environmental quenching
processes such as ram--pressure stripping to act.  


One prediction from this work is that at $z > 0.5$, satellite
quenching times should remain short,  but should scale with the stellar
mass of the satellite and halo mass of the central.  Indeed there is
some evidence for this \citep{balogh16}.   Future studies using
large--area homogeneous datasets probing environments from rich
clusters to the field, combined with the analysis of $N$--body
simulations will enable a detailed study of how the satellite
quenching times depend on these masses and how they evolve with
redshift, which will help address the physics of the environmental
quenching mechanism.  

\acknowledgments  

\editone{The authors thank all their past and current collaborators on
the ZFOURGE survey, whose efforts made this work possible. }  The
authors are grateful to Russell Ryan for sharing his MCMC code in
advance of publication.  The authors also acknowledge fruitful
discussions and conversations with colleagues, especially Avishai
Dekel, Sandra Faber, Steven Finkelstein, and  Yicheng
Guo. \editone{The authors also thank the anonymous referee and
editor, whose comments and suggestions improved the quality and
clarity of this work.}    This work is supported by the National
Science Foundation through grants AST 1410728, 1413317, and 1614668.
\editone{KG acknowledges support for ARC Discovery Projects
DP130101460 and DP130101667.  This paper includes data gathered with
the 6.5 meter Magellan Telescopes located at Las Campanas Observatory,
Chile.  Australian access to the  Magellan Telescopes was supported
through the National Collaborative  Research Infrastructure Strategy
of the Australian Federal Government. }  We acknowledge generous
support from the George P. and Cynthia Woods Institute for Fundamental
Physics and Astronomy at Texas A\&M University.  \editone{This research has made use of NASA's Astrophysics Data System.}


\software{Interactive Data Language (IDL) v8.6,  R \citep{R}}

\appendix

\begin{figure*}
\epsscale{0.9}
\plotone{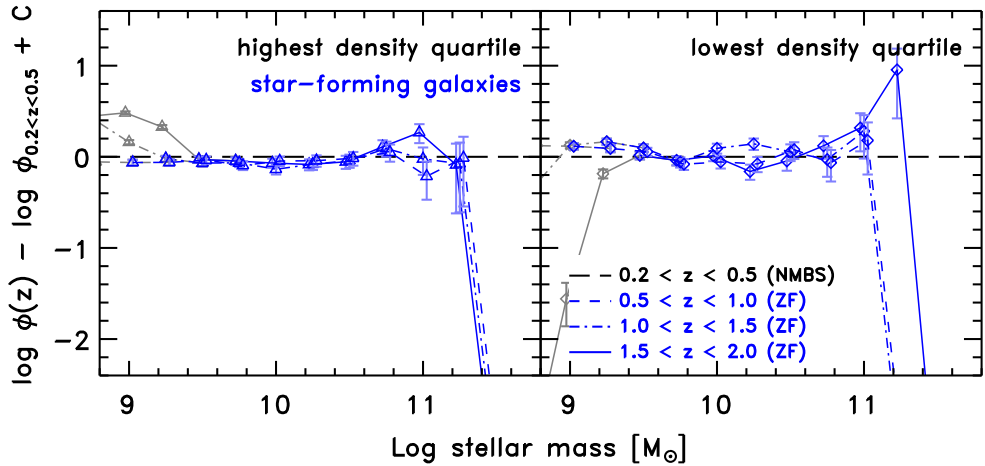}
\epsscale{1.0}
\caption{Relative evolution of the SMF for star-forming galaxies as a
function of environment (similar to Figure~\ref{fig:smf_ratio} above).
The left panel shows the ratio of the SMF in the highest density quartile ($D4$) in each
redshift bin to the SMF of star-forming galaxies in the highest density
quartile at $0.2 < z < 0.5$ (the constant $C$ acts to normalize each
SMF to the same number density at $\log M_\ast/M_\odot$=10.6).    The
right panel shows the same in the lowest
density quartile ($D1$)   The gray-shaded points and lines show data
below the stellar mass completeness limit.  There is no evidence that
the shape of the SMF evolves for quiescent galaxies in the
lowest--density quartile, except possibly at the lowest masses and
highest redshifts where
we observe an excess in the highest density quartiles (and a
corresponding deficit in the lowest density quartiles).  \label{fig:smf_ratio_sf}}
\end{figure*}

\begin{figure*}
\epsscale{0.8}
\plotone{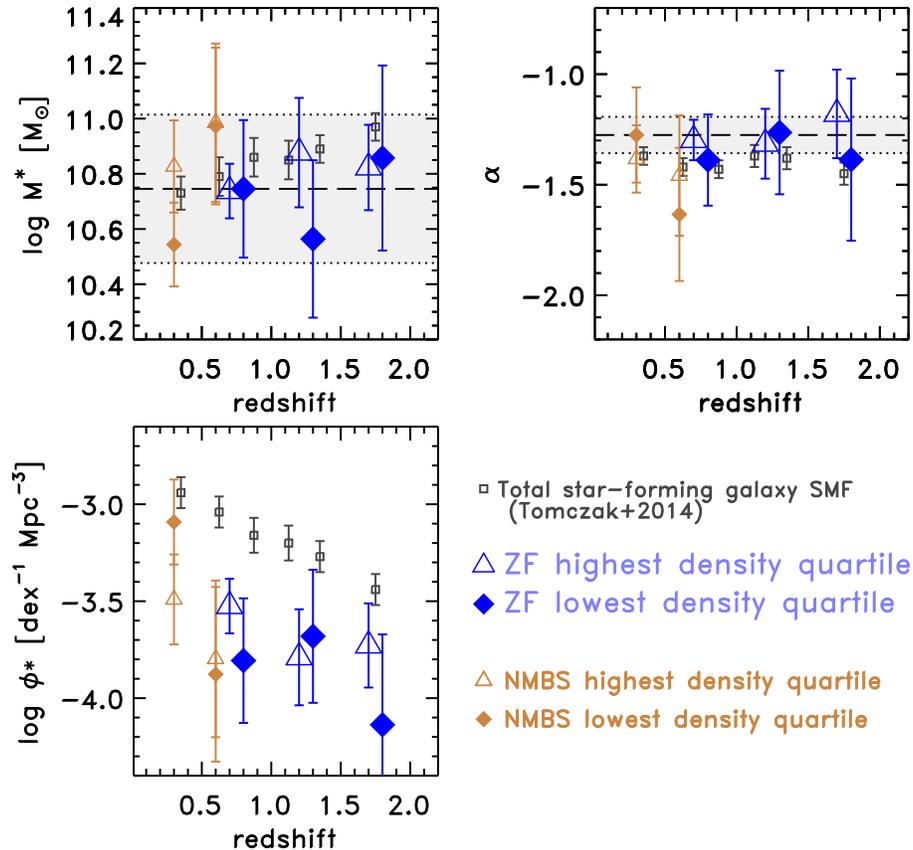}
\epsscale{1.00}
\caption{ Evolution of the  SMF model parameters for star-forming 
galaxies as a function of redshift and galaxy density (similar to
Figure~\ref{fig:smf_parm} above for quiescent galaxies).   The panels
show the evolution with redshift for the characteristic stellar mass
($M^\ast$), the low-mass slope ($\alpha$), and the normalization
($\phi^\ast$)  from the fits to SMF of star-forming galaxies in NMBS and
the ZFOURGE survey.    The symbols correspond to the same SMFs in each
panel, as labeled.  For comparison, the small open, black squares
show the evolution of model parameters for the total SMF of star-forming
galaxies measured by \citet{tomc14}.  \editone{There is no indication that the
values of the characteristic mass, $M^\ast$, or low-mass
slope, $\alpha$ for the low-density environments, evolve with
redshift or environment.}    \label{fig:smf_parm_sf}}
\end{figure*}

\section{Evolution of the SMF of Star-Forming
Galaxies as a Function of Environment}\label{section:appendix}

The emphasis of this \textit{Paper} is on the differential evolution
of the SMF of quiescent galaxies between low--density and
high--density environments.   For completeness, we also discuss here
the differential evolution of the SMF of star-forming galaxies as a
function of environment.    Figure~\ref{fig:smf} shows the SMFs for
star-forming galaxies in the highest ($D4$) and lowest ($D1$) quartiles.

Figures~\ref{fig:smf_ratio} and
\ref{fig:smf_parm} above showed that the change in the shape of the \textit{quiescent}
galaxy SMF results from evolution in the high-density environments.
Here we show similar plots for star-forming galaxies.
Figure~\ref{fig:smf_ratio_sf} shows the ratio of the SMF for
star-forming galaxies in the highest-density quartile and
lowest-density quartile from ZFOURGE at higher redshift compared to
SMF for star-forming galaxies at redshift $0.2 < z < 0.5$ from NMBS.
There is no indication for evolution, with some evidence of an excess
of low-mass galaxies at high galaxy overdensity at high redshifts, and a
deficit of low-mass galaxies at low galaxy overdensity. These occur
below the nominal stellar mass completeness, so we do not attempt to
interpret them in great detail.  However, if this is real, then it would
suggest that lower--mass star-forming galaxies are preferentially
found in regions of higher overdensity, but only at the highest
redshifts ($z \gg 1$).  

\editone{To quantify the differences in the SMF of star-forming
  galaxies, we have applied the non-parametric KS and MWW statistical tests to the
  distribution of galaxy stellar masses in the different subsamples,
  similar to the analysis of quiescent galaxies in
  \S~\ref{section:smf}.     Table~\ref{table:pvalues} presents the
  results ($p$--values) of these tests comparing the different stellar--mass
  distributions. }

\editone{ We compared the unbinned stellar--mass distributions of
star-forming galaxies in the highest environmental density quartile
($D4$) at $0.5 < z < 1.0$ to the stellar mass distributions of
quiescent galaxies in $D4$ at higher redshift.   Comparing the
stellar--mass distribution of the $0.5 < z < 1.0$ galaxies to that at
$1.0 < z < 1.5$, the KS test gives $p=0.074$ and the MWW test gives
$p=0.27$.  This could indicate very weak evidence that the
distributions are drawn from different parent distributions, which may
indicate weak evolution \citep[and consistent with evolution in the
shape of the total SMF reported for star-forming galaxies,
see,][]{tomc14}.  However, comparing the stellar--mass distribution of
the $0.5 < z < 1.0$ star-forming galaxies to those at $1.5 < z < 2.0$,
the KS test and MWW test show no evidence they are drawn from
different parent populations, with $p$--values of 0.23 and 0.82,
respectively. (Comparing the $1.0 < z < 1.5$ with the $1.5 < z < 2.0$
populations shows no evidence for differences either).   } 

\editone{Similarly, the stellar--mass distributions of star-forming
  galaxies in the lowest environmental density quartile ($D1$) show no
  convincing evidence they change with redshift.  Comparing the
  stellar--mass distribution of the $0.5 < z < 1.0$ galaxies to that
  at $1.0 < z < 1.5$, the KS test gives $p=0.027$ and the MWW test
  gives $p=0.30$.  Comparing the stellar--mass distribution of the
  $0.5 < z < 1.0$ galaxies to that $1.5 < z < 2.0$, the KS test gives
  $p=0.12$ and the MWW test gives $p=0.74$ (with no additional
  insight by comparing the $1.0 < z < 1.5$ galaxies to those at $1.5 <
  z < 2.0$).    Again, there is no strong evidence that the
  distribution of stellar mass is evolving strongly (with the possible
  exception for evolution at $z < 1.0$ in both the $D1$ and $D4$
  populations, but the evidence is weak). }

As with the quiescent galaxy SMF, we also fit Schechter functions
to the star-forming galaxy SMFs at each redshift and overdensity.
Figure~\ref{fig:smf_parm_sf} shows the evolution of the values of
the characteristic mass, $M^\ast$, low-mass slope, $\alpha$, and
normalization, $\phi^\ast$, for the star-forming SMFs as a function of
redshift and overdensity.     In contrast to quiescent galaxies (cf.\
Figure~\ref{fig:smf_parm}), there is no evidence of evolution in the
shape of the SMF of star-forming galaxies over the stellar--mass range
and redshift range studied here:  the values of $M^\ast$ or $\alpha$
are consistent across redshift and overdensity for
the SMF of star-forming galaxies.

\bibliographystyle{yahapj}
\bibliography{alpharefs}{}

\end{document}